\documentclass[dvips]{acta}
\usepackage{supertabular,lscape,epsfig}
\usepackage{amssymb}
\usepackage{amsmath}
\DeclareSymbolFont{ppa}{OT1}{ppl}{m}{it}
\DeclareMathSymbol{\vv}{\mathalpha}{ppa}{'166}

\newfont{\hb}{rphvb at 10pt}
\newfont{\hbo}{rphvbo at 10pt}
\newfont{\bitt}{rptmbi at 12pt}
\newfont{\bits}{rptmbi at 11pt}

\SetPages{323}{338}

\SetVol{63}{2013}

\begin{document}

\newcommand{\TabApp}[2]{\begin{center}\parbox[t]{#1}{\centerline{
  {\bf Appendix}}
  \vskip2mm
  \centerline{\small {\spaceskip 2pt plus 1pt minus 1pt T a b l e}
  \refstepcounter{table}\thetable}
  \vskip2mm
  \centerline{\footnotesize #2}}
  \vskip3mm
\end{center}}

\newcommand{\TabCapp}[2]{\begin{center}\parbox[t]{#1}{\centerline{
  \small {\spaceskip 2pt plus 1pt minus 1pt T a b l e}
  \refstepcounter{table}\thetable}
  \vskip2mm
  \centerline{\footnotesize #2}}
  \vskip3mm
\end{center}}

\newcommand{\TTabCap}[3]{\begin{center}\parbox[t]{#1}{\centerline{
  \small {\spaceskip 2pt plus 1pt minus 1pt T a b l e}
  \refstepcounter{table}\thetable}
  \vskip2mm
  \centerline{\footnotesize #2}
  \centerline{\footnotesize #3}}
  \vskip1mm
\end{center}}

\newcommand{\MakeTableApp}[4]{\begin{table}[p]\TabApp{#2}{#3}
  \begin{center} \TableFont \begin{tabular}{#1} #4 
  \end{tabular}\end{center}\end{table}}

\newcommand{\MakeTableSepp}[4]{\begin{table}[p]\TabCapp{#2}{#3}
  \begin{center} \TableFont \begin{tabular}{#1} #4 
  \end{tabular}\end{center}\end{table}}

\newcommand{\MakeTableee}[4]{\begin{table}[htb]\TabCapp{#2}{#3}
  \begin{center} \TableFont \begin{tabular}{#1} #4
  \end{tabular}\end{center}\end{table}}

\newcommand{\MakeTablee}[5]{\begin{table}[htb]\TTabCap{#2}{#3}{#4}
  \begin{center} \TableFont \begin{tabular}{#1} #5 
  \end{tabular}\end{center}\end{table}}

\newfont{\bb}{ptmbi8t at 12pt}
\newfont{\bbb}{cmbxti10}
\newfont{\bbbb}{cmbxti10 at 9pt}
\newcommand{\uprule}{\rule{0pt}{2.5ex}}
\newcommand{\douprule}{\rule[-2ex]{0pt}{4.5ex}}
\newcommand{\dorule}{\rule[-2ex]{0pt}{2ex}}
\def\thefootnote{\fnsymbol{footnote}}
\begin{Titlepage}
\Title{Eclipsing Binary Stars in the OGLE-III Fields\\ of the Small
Magellanic Cloud\footnote{Based on observations obtained with the 1.3-m
Warsaw telescope at the Las Campanas Observatory of the Carnegie
Institution for Science.}}
\Author{M.~~P~a~w~l~a~k$^1$,~~ D.~~G~r~a~c~z~y~k$^2$,~~ I.~~S~o~s~z~y~ñ~s~k~i$^1$,~~ P.~~P~i~e~t~r~u~k~o~w~i~c~z$^1$,\\
R.~~P~o~l~e~s~k~i$^{1,3}$,~~ A.~~U~d~a~l~s~k~i$^1$,~~ M.\,K.~~S~z~y~m~a~ñ~s~k~i$^1$,~~ M.~~K~u~b~i~a~k$^1$,\\
G.~~P~i~e~t~r~z~y~ñ~s~k~i$^{1,2}$,~~
£.~~W~y~r~z~y~k~o~w~s~k~i$^{1,4}$,~~
K.~~U~l~a~c~z~y~k$^1$,~~ S.~~K~o~z~³~o~w~s~k~i$^1$
~~and~~ J.~~S~k~o~w~r~o~n$^1$}
{$^1$Warsaw University Observatory, Al. Ujazdowskie 4, 00-478 Warszawa, Poland\\
e-mail: (mpawlak,soszynsk,pietruk,rpoleski,udalski)@astrouw.edu.pl\\
$^2$Departamento de Astronomia, Universidad de Concepción, Casilla 160-C, Chile\\
e-mail: darek@astro-udec.cl\\
$^3$Department of Astronomy, The Ohio State University, 140 West 18th Avenue, Columbus, OH 43210, USA\\
$^4$Institute of Astronomy, University of Cambridge, Madingley Road, Cambridge CB3 0HA, UK
}
\Received{October 2, 2013}
\end{Titlepage}

\Abstract{We present a large sample of eclipsing binary stars detected in
the Small Magellanic Cloud fields covering about 14 square degrees that
have been monitored for eight years during the third phase of the OGLE
survey. This is the largest set of such variables containing 6138 objects,
of which 777 are contact and 5361 non-contact binaries. The estimated
completeness of this sample is around 82\%.

We analyze the statistical properties of the sample and present
selected interesting objects: 32 systems having eccentric orbit with
visible apsidal motion, one Transient Eclipsing Binary, ten RS~CVn type
stars, 22 still unexplained Double-Periodic Variable stars, and 15
candidates for doubly eclipsing quadruple systems. Based on the OGLE-III
proper motions, we classified 47 binaries from our sample as foreground
Galactic stars. We also list candidates suitable for the SMC distance
determination.}{binaries: eclipsing -- Stars: variables: general --
Magellanic Clouds}
   
\Section{Introduction} 
Eclipsing binary stars are very important objects in astrophysical studies,
as they allow us to measure basic physical parameters of their components,
which are often difficult or impossible to obtain for a single object. For
instance, for the first time, the dynamical mass of a classical Cepheid
being a component of a wide eclipsing system has recently been accurately
determined by Pietrzyñski \etal (2010). Eclipsing binaries, beside
astrometric binaries, are the only source of the very accurate stellar mass
determinations (\eg Torres \etal 2009, He³miniak \etal 2009, Zola \etal
2010, Deb and Singh 2011). Moreover, they can be used for testing pulsation
and evolutionary models (Torres \etal 2010). Recent discoveries show that
very specific types of stars, like Binary Evolution Pulsating stars
(Pietrzyñski \etal 2012, Maxted \etal 2013), can form in binary systems.

Eclipsing binaries are also a powerful tool for obtaining distance
measurements. The most accurate determination of the distance to the Large
Magellanic Cloud (LMC) using late type eclipsing binaries (Pietrzyñski
\etal 2013) can be the best example here. A somewhat less accurate method,
based on early type binaries allowed measurements of distances to M31
(Valls-Gabaud 2013), M33 (Guinan \etal 2013), and several other Local Group
galaxies (Bonanos 2013). Moreover, eclipsing binary stars can also be used
to determine the age of star clusters (\eg Grundahl \etal 2008, Meibom
\etal 2009, Kaluzny \etal 2013). Although binary systems are believed to be
quite common, only a few percent of them can be observed as eclipsing
(S{\"o}derhjelm and Dischler 2005).

The first confirmed eclipsing systems in the Small Magellanic Cloud (SMC)
were presented by Shapley and Nail (1953) in their list of fifty binary
stars in the fields of the Magellanic Clouds. The catalog of variable stars
in the SMC by Payne-Gaposchkin and Gaposchkin (1966) contains 34 eclipsing
binaries.  In the 1980s, binaries in the SMC area were investigated in
Breysacher \etal (1982) and Davidget (1988). More recent studies based on
microlensing surveys data resulted in the discovery of much larger number
of these objects. Udalski \etal (1998) and Wyrzykowski \etal (2004)
detected 1459 and 1351 eclipsing binaries, respectively, in the OGLE-II
databases. The two independent searches brought a total number of 1914
stars. The MACHO catalog by Faccioli \etal (2007) contains 1509 objects, of
which 698 were matched with the OGLE data.

In this paper we present a new set of 6138 eclipsing binary stars in the
Small Magellanic Cloud from the third phase of the Optical Gravitational
Lensing Experiment (OGLE-III). It constitutes the next part of the OGLE-III
Catalog of Variable Stars (OIII-CVS). This is the third part of OIII-CVS
dedicated to binary stars. Two previous parts were devoted to the LMC
(Graczyk \etal 2011) and the Galactic disk fields (Pietrukowicz \etal
2013).

In the following sections of this paper, we describe: observations and
their reductions (Section~2), selection and classification of the eclipsing
stars (Section~3), catalog of eclipsing binaries in the SMC (Section~4),
and paricularly interesting objects found (Section~5). In the last section
(Section~6), we summarize our results.

\Section{Observations and Data Reduction}
All the data presented in this paper were collected with the 1.3-m Warsaw
telescope at the Las Campanas Observatory in Chile. The observatory is
operated by the Carnegie Institution for Science. During the OGLE-III
phase, the telescope was equipped with a mosaic eight-chip camera, with the
field of view of about $35\arcm\times35\arcm$ and the scale of
0\zdot\arcs26~pixel$^{-1}$. For details of the instrumentation setup we
refer to Udalski (2003).

Altogether 41 OGLE-III fields covering about 14 square degrees in the sky
were observed toward the SMC and about 6~million sources were detected.
Approximately 730 photometric points per star were secured over a timespan
of eight years, between July 2001 and May 2009. About 90\% of observations
were taken in the standard {\it I}-band, while the remaining measurements
were taken in the {\it V}-band. The OGLE data reduction pipeline (Udalski
2003) is based on the Difference Image Analysis technique (Alard and Lupton
1998, Wo¼niak 2000). A full description of the reduction techniques,
photometric calibration and astrometric transformations can be found in
Udalski \etal (2008).

\Section{Selection and Classification}
Two different methods were applied to identify eclipsing binaries. First,
the classification was performed with machine learning algorithms. With
this approach, the program is ``learning'' how to distinguish a light curve
of a binary system based on already classified examples. It produces a
classifier, in this case a decision tree, which is used to perform
classification of new objects. Three different algorithms: C4.5, Naive
Bayes Tree (NBT), and Logistic Model Tree (LMT) were used to construct the
tree. The C4.5 algorithm splits the set of training examples into subsets
in each node of the tree using entropy decrease as a criterion for choosing
the optimal parameter for splitting. In the NBT algorithm the Bayes'
theorem is used for splitting in the nodes and in the LMT -- a logistic
function. The construction of the tree continues recurrently until it gives
uniform subsets or reaches stop criteria like the minimum number of
elements in a subset necessary for splitting. For the details of the C4.5,
NBT, and LMT algorithms we refer to Quinlan (1993), Kohavi (1996), and
Landwehr \etal (2005), respectively.

In our search, the training set was constructed based on 26\,122 eclipsing
binaries from the LMC OGLE-III set of these objects (Graczyk \etal 2011)
and additional 135\,602 random non-binary objects from the LMC. The
following statistical parameters: standard deviation, skewness, and
kurtosis of the brightness were used for the parametrization of the light
curves of stars from the training set and for objects which were to be
classified. Only the stars which were identified as eclipsing binaries with
all three algorithms were considered as positive detections. Such an
approach was used to reduce the number of false detections. Implementation
of the decision tree algorithms used in this study comes from the data
mining software {\sc Weka} (Hall \etal 2009).

The second approach we adopted was with the algorithm used by Graczyk \etal
(2011) for the LMC fields. This method is also based on a statistical
parametrization of the light curve. For details of the method we refer the
reader to Graczyk and Eyer (2010).

Finally, our set of eclipsing binaries was supplemented for completeness
with the eclipsing stars found when preparing other parts of the OIII-CVS
by Soszyñski \etal (2010ab, 2011) and missing stars from the OGLE-II
catalogs (Udalski \etal 1998, Wyrzykowski \etal 2004) and the MACHO catalog
(Faccioli \etal 2007).

All light curves were subject to visual inspection. The final list of
eclipsing binaries contains 6138 objects. They were divided into two types:
contact binaries (EC) and detached and semi-detached binaries (non-EC). The
classification was based on the shape of the light curves. We found 777 EC
and 5361 non-EC systems.

Out of 4822 stars brighter than $I=19$~mag in our final set, 3375 were
detected with the machine learning method and 4308 with the traditional
approach. This gives relative effectiveness of the methods at 70\% and
89\%, respectively. Thus the search for eclipsing binaries can be
successfully performed with the machine learning algorithms, although the
traditional method still gives better results. A more detailed
parametrization of the light curves could likely improve the final
performance.

\Section{Catalog of Eclipsing Binary Stars in the SMC}  
The results of our search for eclipsing binaries in the SMC are presented
in the form of the catalog. The catalog containing table of objects, their
parameters, cross-identification with MACHO and General Catalog of Variable
Stars (GCVS, Samus \etal 2011), {\it I}- and {\it V}-band photometry, and
finding charts, is available to the astronomical community from the OGLE
Internet archive:

\begin{center}
{\it http://ogle.astrouw.edu.pl/}\\
{\it ftp://ftp.astrouw.edu.pl/ogle/ogle3/OIII-CVS/smc/ecl/}\\
\end{center}

The FTP site is organized as follows. The list of eclipsing binaries with
their J2000 equatorial coordinates, classification, identifications in the
OGLE-II and OGLE-III databases, MACHO and GCVS are given in the {\sf
ident.dat} file. The stars are arranged in order of increasing right
ascension and designated as OGLE-SMC-ECL-NNNN, where NNNN is a four-digit
consecutive number.

The data table containing the following observational parameters of each
object: maximum {\it I}- and {\it V}-band magnitude, orbital period,
amplitude, epoch of the main eclipse, standard deviation, skewness, and
kurtosis of the measurements is given in the file {\sf ecl.dat}. The
time-series {\it I}- and {\it V}-band photometry are given in separate
files in the subdirectory {\sf phot/}. The subdirectory {\sf fcharts/}
contains finding charts for all objects. These are $60\arcs\times60\arcs$
subframes of the {\it I}-band DIA reference images, oriented with North up,
and East to the left. The file {\sf remarks.txt} contains additional
information about interesting objects. Positions in the sky of all
identified objects are shown in Fig.~1.

Periods were determined using the {\sc Fnpeaks} code kindly provided by
Z. Ko\-³aczkowski and the {\sc AoV} code by Schwarzenberg-Czerny
(1996). For most objects, they were corrected with the {\sc Tatry}
algorithm (Schwarzenberg-Czerny 1996). The maximum of brightness was set by
fitting the Fourier series for sinusoidal-like curves. For non-sinusoidal
(detached) ones it was computed as the median minus the standard deviation
of the magnitude. For the cases when both approaches failed, the maximum
brightness was set manually.
\begin{figure}[htb]
\centerline{\includegraphics[width=125mm]{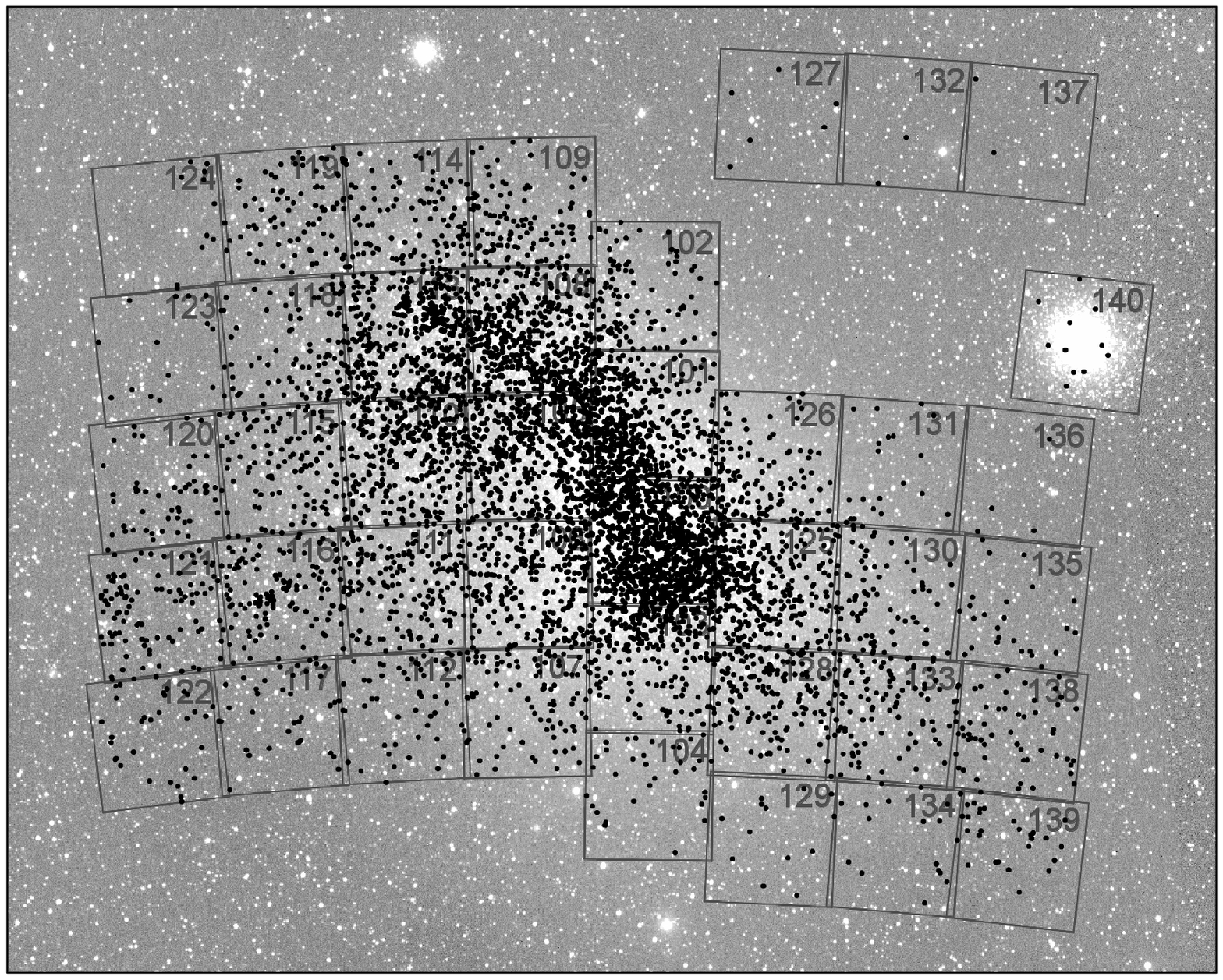}}
\vskip5pt
\FigCap{Positions of identified eclipsing binary systems from our
catalog. Black contours represent the OGLE-III fields. The image of the SMC in
the background comes from the ASAS project (Pojmañski 1997).}
\end{figure}

The cross-match with the MACHO catalog allowed us to evaluate the
completeness of our set. Out of 1509 objects found by MACHO, 1253 were
identified independently. We identified 241 of the missing stars in the
OGLE database, although they were not found while searching for
binaries. These objects were then added to the catalog for completeness.
Estimated completeness of our sample, based on this comparison is:
$1253/(1253+241)\approx83\%$.

Another way of evaluating the completeness of our sample is by using the
stars located in the overlapping regions of the observed fields. We found
273 such objects, of which 188 were detected twice, independently in each
field. This gives us a completeness of: $2x/(1+x)\approx82\%$, where $x
=188/273$. However, these estimations correspond to the bright eclipsing
stars with high amplitudes.  For faint and low-amplitude binaries the
completeness is significantly lower.

The stars from the catalog have {\it I}-band brightness in the range
$12.5<I<20.5$~mag. In Fig.~2, we present the brightness distribution. It
reaches the maximum around 18.5~mag. Then the completeness of the catalog
drops significantly. Amplitude distribution is shown in Fig.~3. The
majority of systems have amplitudes between 0.2~mag and 0.4~mag.
\begin{figure}[htb]
\centerline{\includegraphics[angle=270, width=10cm, bb=110 50 505 765]{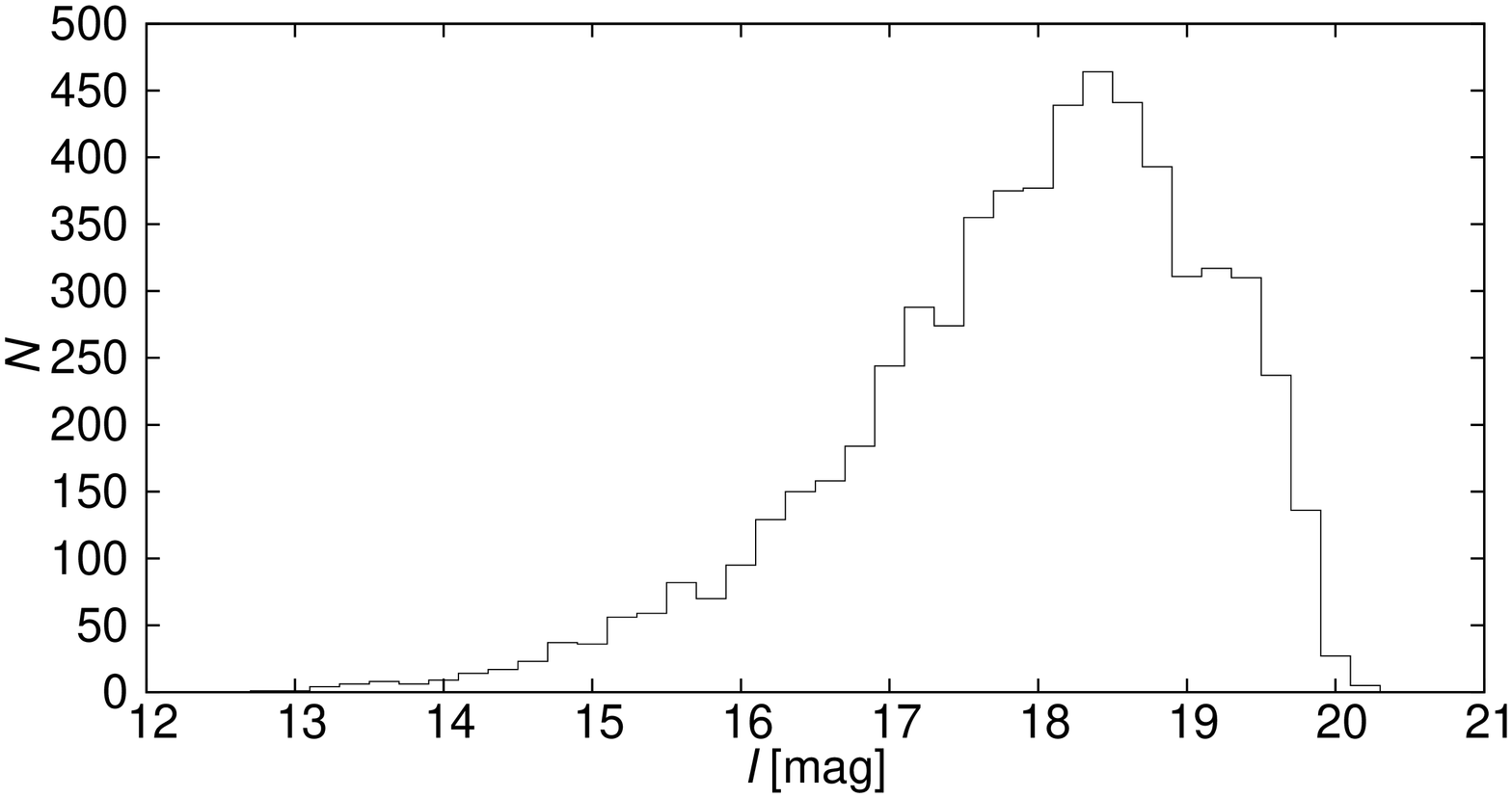}}
\FigCap{{\it I}-band magnitude distribution for eclipsing binaries from the
OGLE-III SMC set.}
\centerline{\includegraphics[angle=270, width=10cm, bb=110 50 505 765]{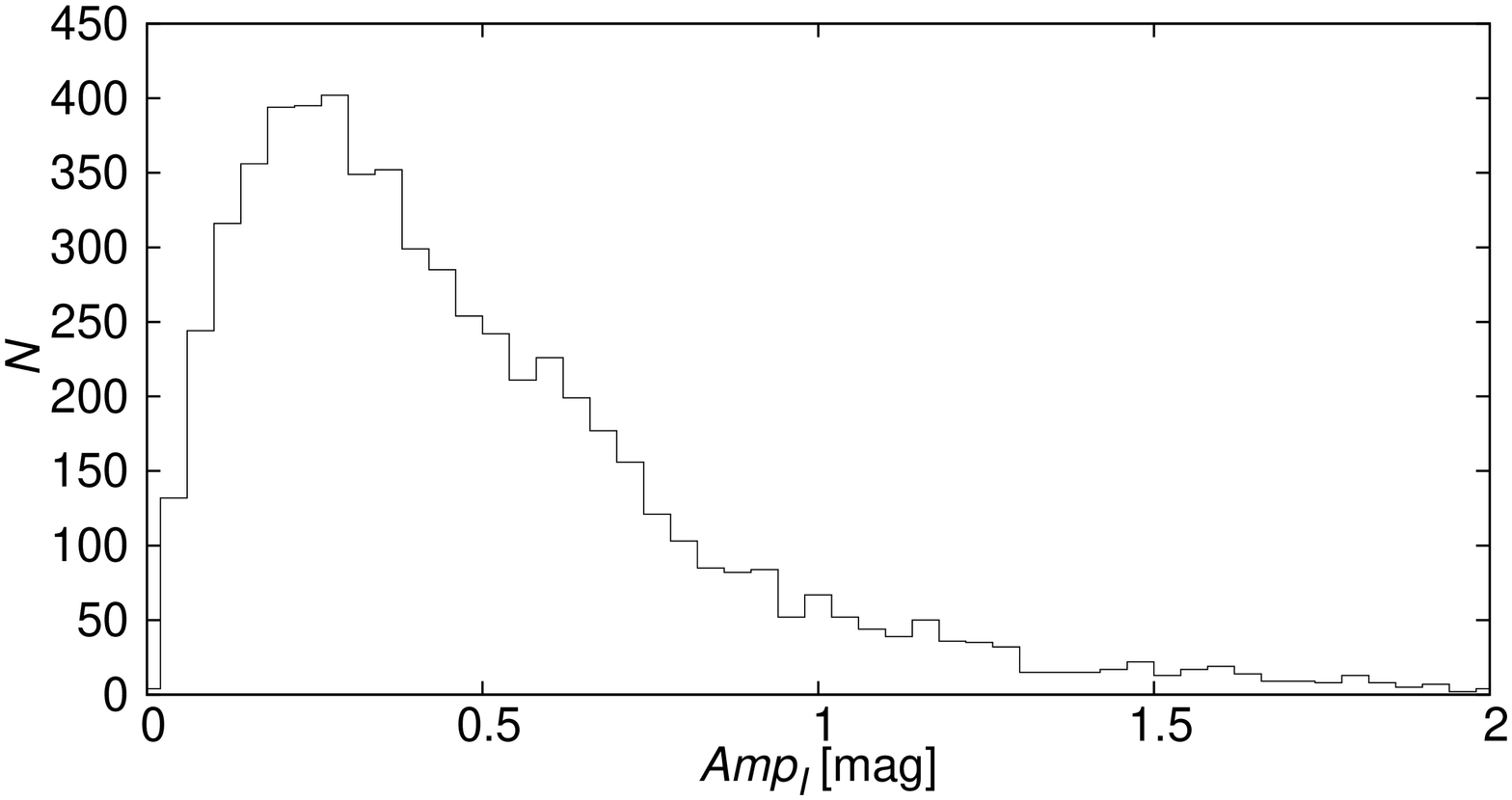}}
\FigCap{Amplitude distribution for eclipsing binaries from the 
OGLE-III SMC set.}
\end{figure}

In Fig.~4, we present the distribution of the orbital periods of our
eclipsing systems. They are divided into EC and non-EC, though the
classification may be uncertain in some cases as it was based on the visual
analysis of the light curves only. For periods shorter than 0.6~d the
number of EC and non-EC systems is similar. For longer periods the vast
majority of the systems are non-EC ones. The shortest period in the sample
is 0.25882~d and it was found for a contact system --
OGLE-SMC-ECL-2992. The longest period is 1220~d for OGLE-SMC-ECL-1637.

\begin{figure}[h]
\centerline{\includegraphics[angle=270, width=10cm, bb=110 50 505 765]{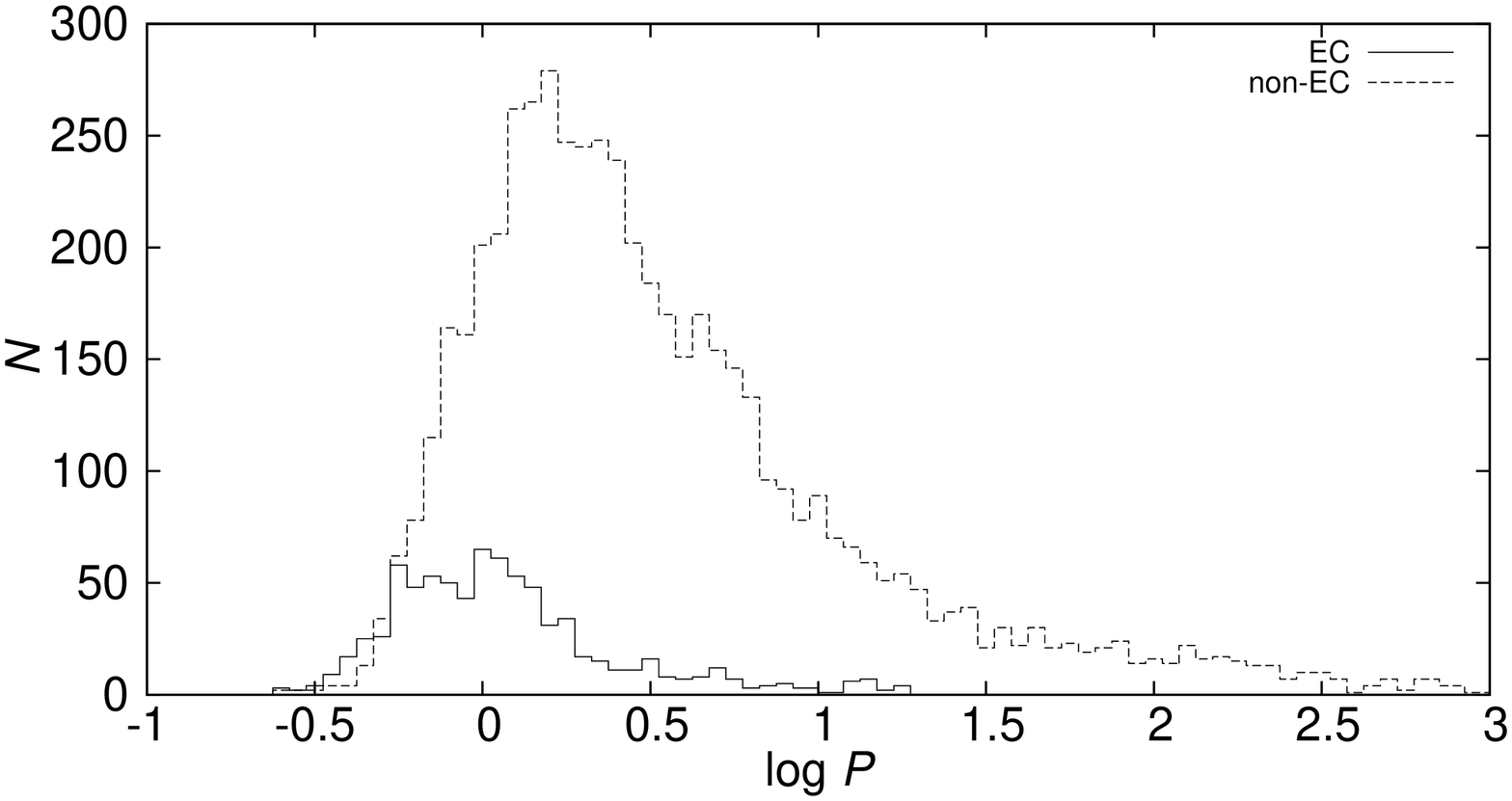}}
\FigCap{Period distribution for contact (solid line) and non-contact (dashed line)
eclipsing binaries from the OGLE-III SMC set.}
\centerline{\includegraphics[angle=270,width=139mm]{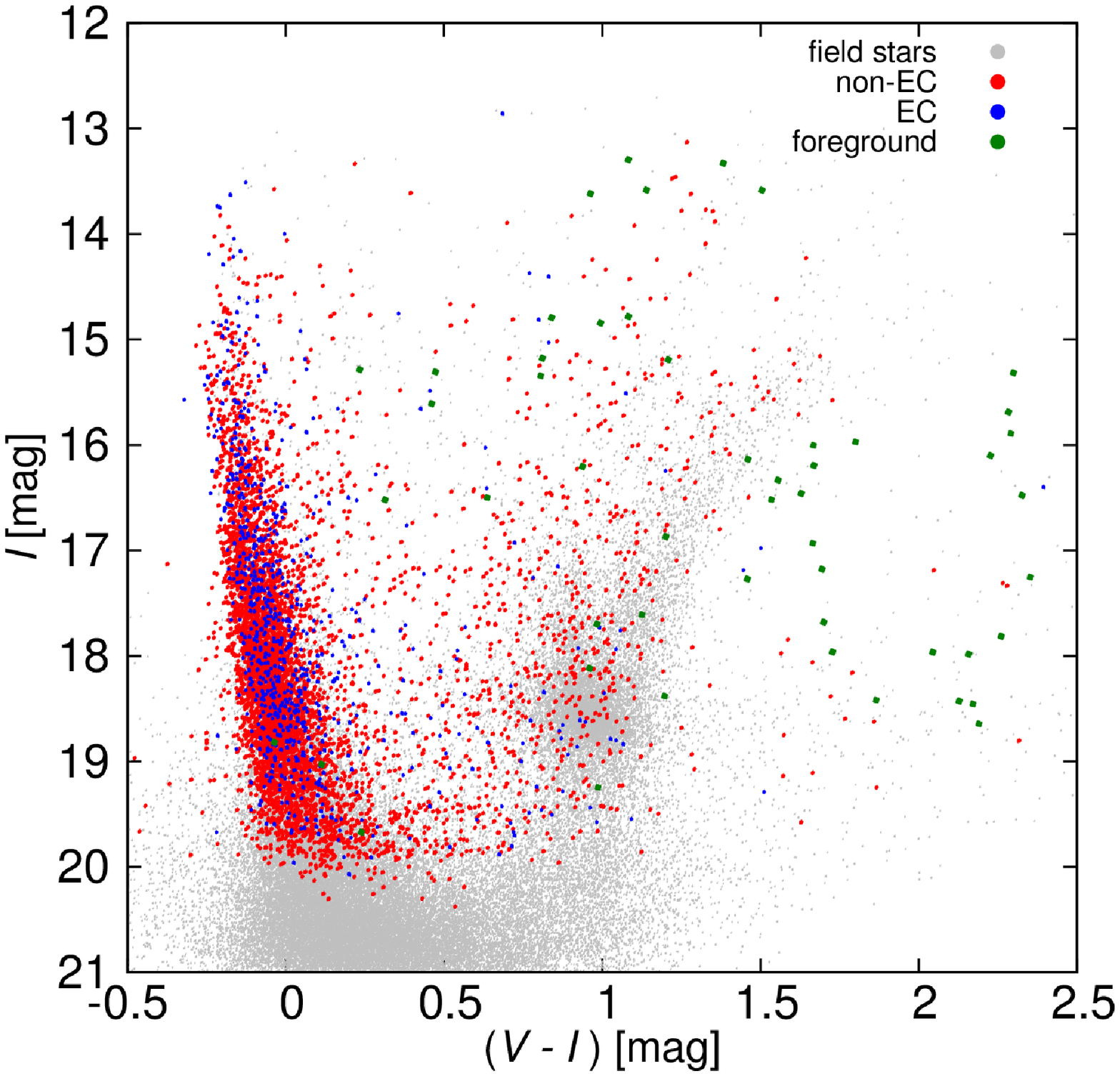}}
\FigCap{Color--magnitude diagram for the eclipsing binaries from the 
OGLE-III SMC set.}
\end{figure}
We analyzed the proper motions of the stars in our set (Poleski \etal 2012)
to determine if they were located either in the SMC or in the halo of the
Galaxy. Only 47 objects reveal high proper motion ($>6$~mas/yr) and they
were classified as foreground stars. This constitutes less than 1\% of our
sample. Six of them are EC and 41 are non-EC systems, which gives a type
ratio comparable to the rest of the binaries. These objects are marked in
the {\sf remarks.txt} file in the OGLE archive.

Color--magnitude diagram of our eclipsing binaries is presented in
Fig.~5. It is worth noticing that EC binaries are located mostly on the main
sequence, while non-EC ones populate both -- the main sequence and the red
giant branch. Foreground objects were also marked on the diagram. Most of
them are red dwarf stars from the Galaxy.

\Section{Interesting Objects in the OGLE-III Set of Eclipsing Binaries}  
\subsection{Candidates for Doubly Eclipsing Quadruple Systems}
In our set of eclipsing binaries, we identified 15 single objects showing
variability from more than one eclipsing system. For each of them we
subtracted the main period $P_1$ to obtain the second period $P_2$.
Examples of light curves with separated components are shown in Fig.~6.
The full list of candidates is presented in Table~1.
\MakeTable{lcr}{12.5cm}{List of candidates for doubly eclipsing quadruple systems}
{\hline
\noalign{\vskip3pt}
\multicolumn{1}{c}{ID} & $P_1$ [d] & \multicolumn{1}{c}{$P_2$ [d]} \\
\noalign{\vskip3pt}
\hline
\noalign{\vskip3pt}
OGLE-SMC-ECL-0629 & 3.95327 & 244.79804 \\
OGLE-SMC-ECL-1076 & 6.40349 &   4.30215 \\
OGLE-SMC-ECL-1758 & 0.92917 &   3.73518 \\
OGLE-SMC-ECL-2036 & 1.25371 &  21.75096 \\
OGLE-SMC-ECL-2141 & 0.56554 &   1.27330 \\
OGLE-SMC-ECL-2208 & 5.72602 &   2.61777 \\
OGLE-SMC-ECL-2529 & 1.07455 &   6.54472 \\
OGLE-SMC-ECL-2586 & 1.25169 &   1.51224 \\
OGLE-SMC-ECL-2715 & 0.76321 &   1.02086 \\
OGLE-SMC-ECL-2896 & 0.65978 &   1.18166 \\
OGLE-SMC-ECL-3284 & 1.01122 &   2.43480 \\
OGLE-SMC-ECL-4418 & 0.71821 &   3.26509 \\
OGLE-SMC-ECL-4731 & 0.73811 &   0.61356 \\
OGLE-SMC-ECL-4908 & 2.55792 &   2.85180 \\
OGLE-SMC-ECL-5015 & 0.76283 &   1.15616 \\
\noalign{\vskip3pt}
\hline}

For each of the candidates, we analyzed changes of the centroid positions
(\cf Pietrukowicz \etal 2013). We did not find any correlation with the
phase of the periods $P_1$ and $P_2$. This suggests that the objects are
rather physically bound quadruple systems than two blended binaries.
However, additional observations, including spectroscopic measurements, are
required to provide a definitive conclusion.

\begin{figure}[p]
OGLE-SMC-ECL-2208\\
\includegraphics[angle=270,width=60mm]{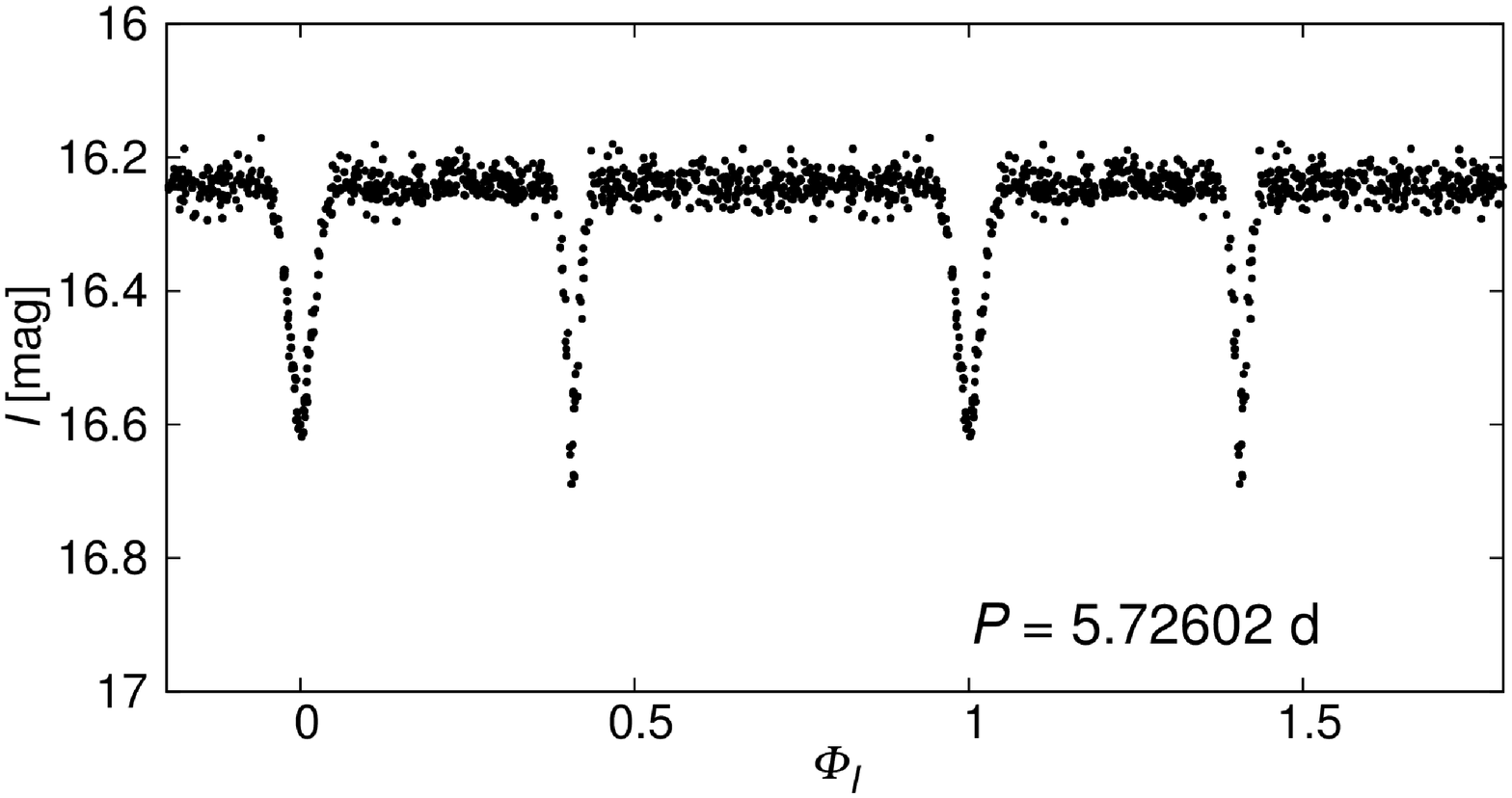}\hfill \includegraphics[angle=270,width=60mm]{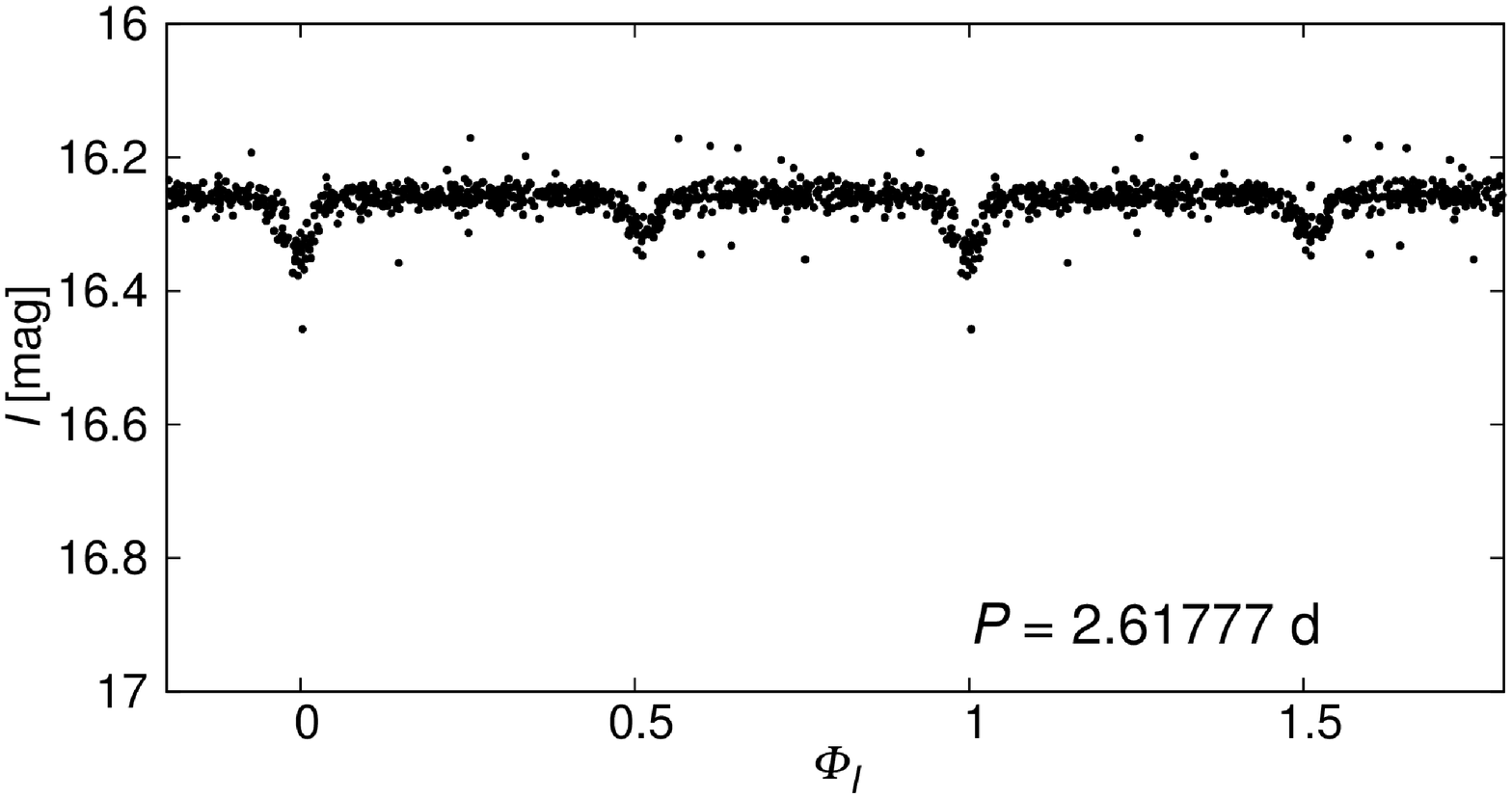} \\
OGLE-SMC-ECL-2715\\
\includegraphics[angle=270,width=60mm]{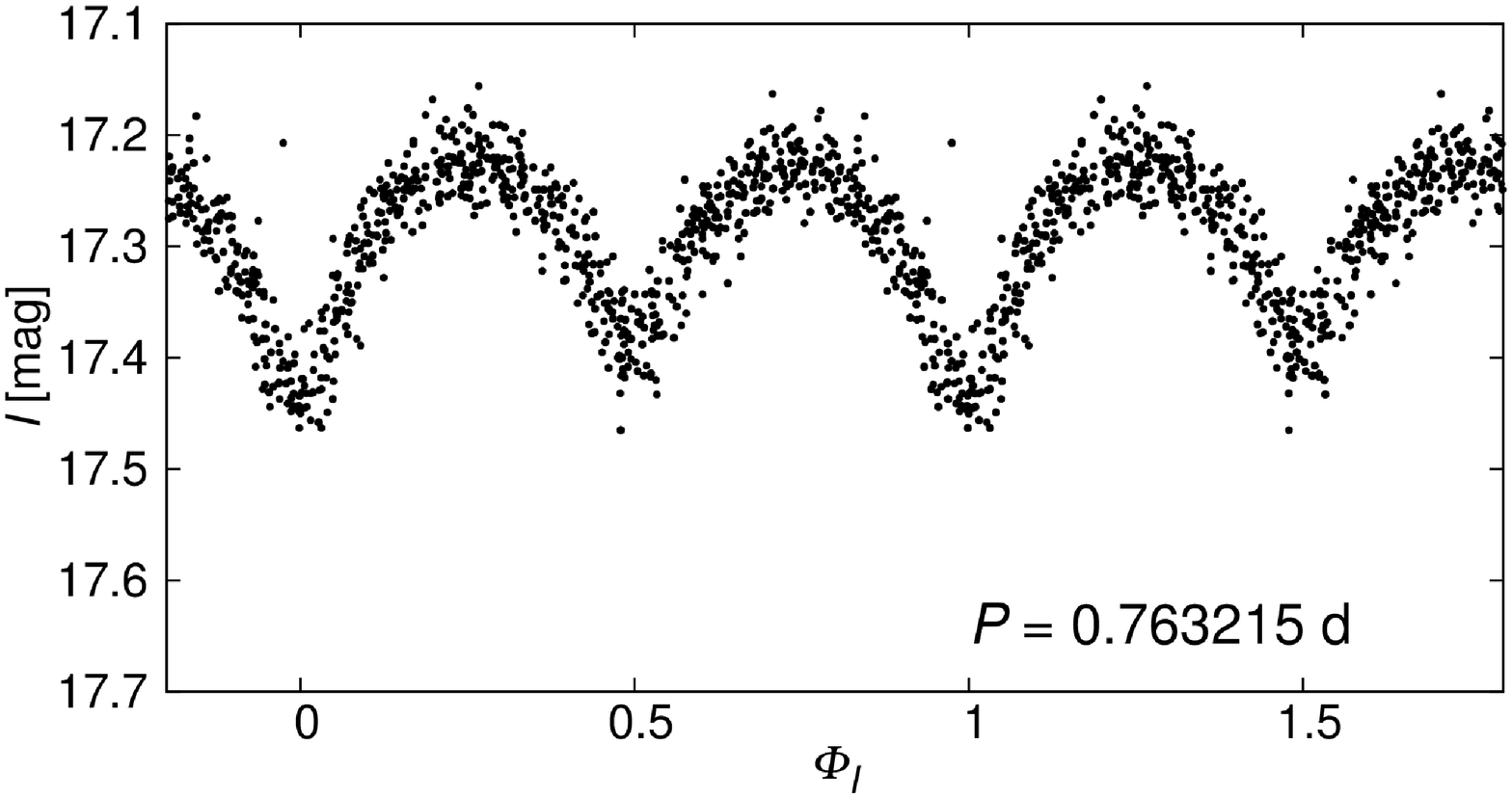}\hfill \includegraphics[angle=270,width=60mm]{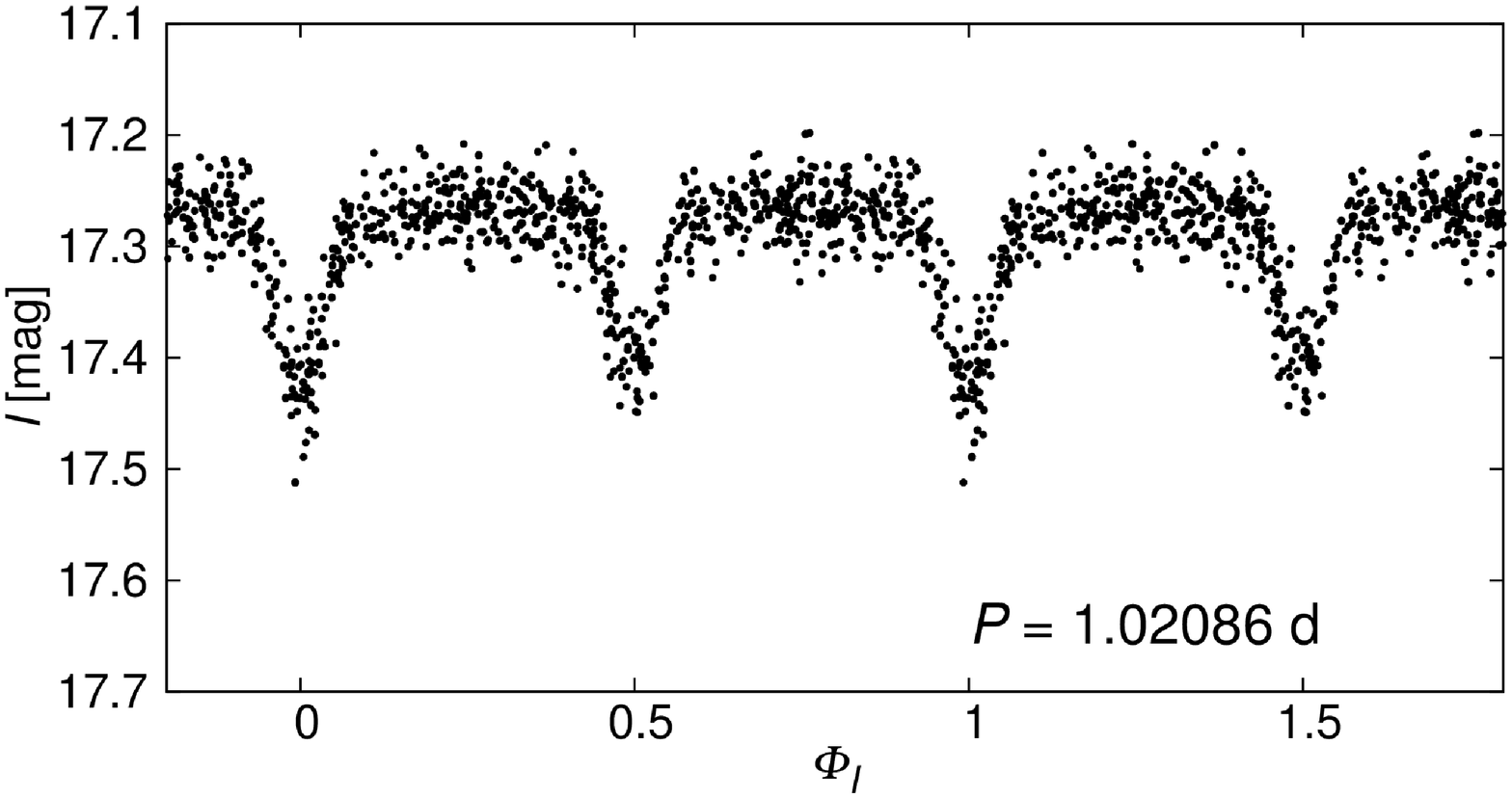} \\
OGLE-SMC-ECL-4418\\
\includegraphics[angle=270,width=60mm]{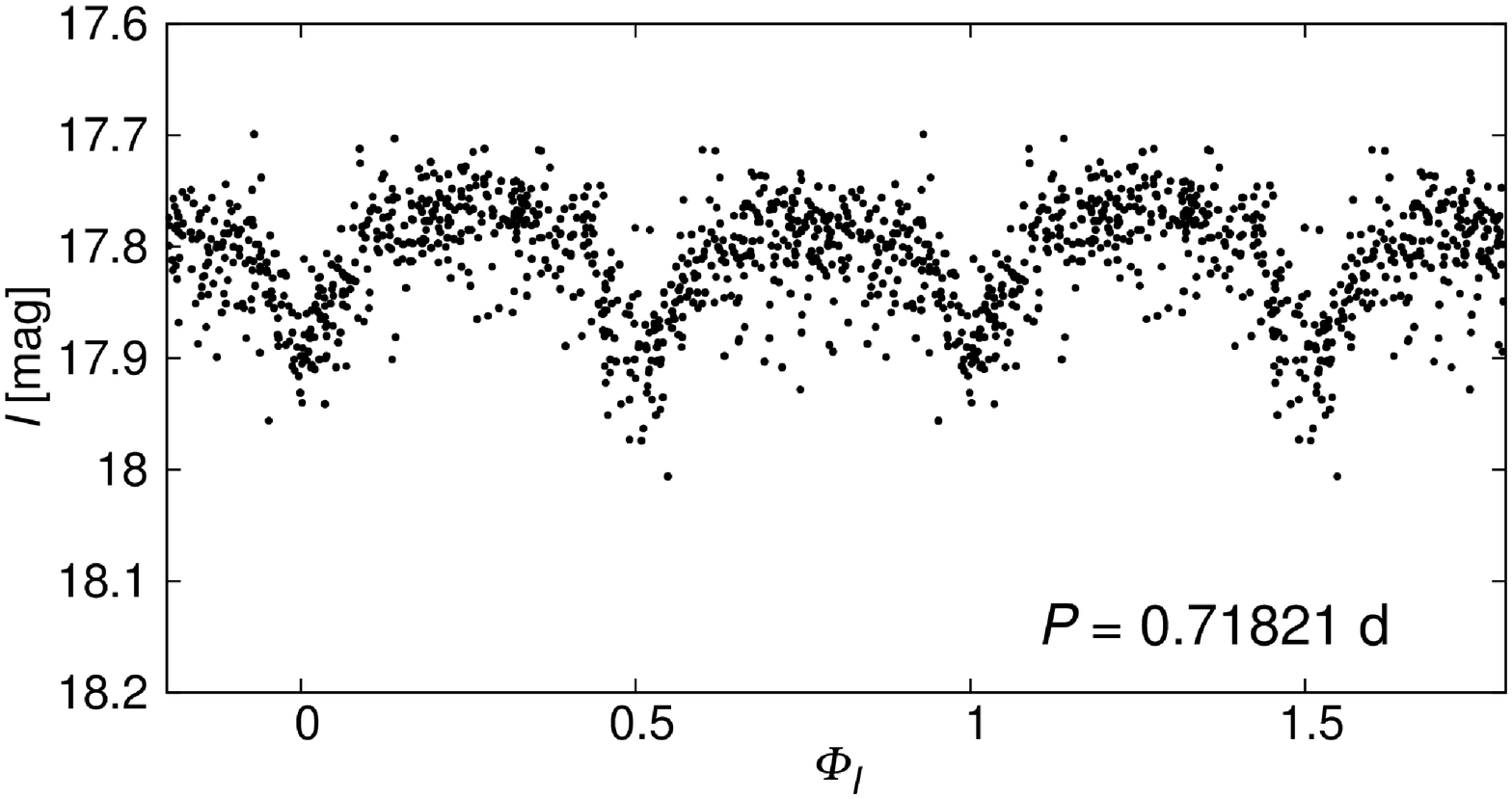}\hfill \includegraphics[angle=270,width=60mm]{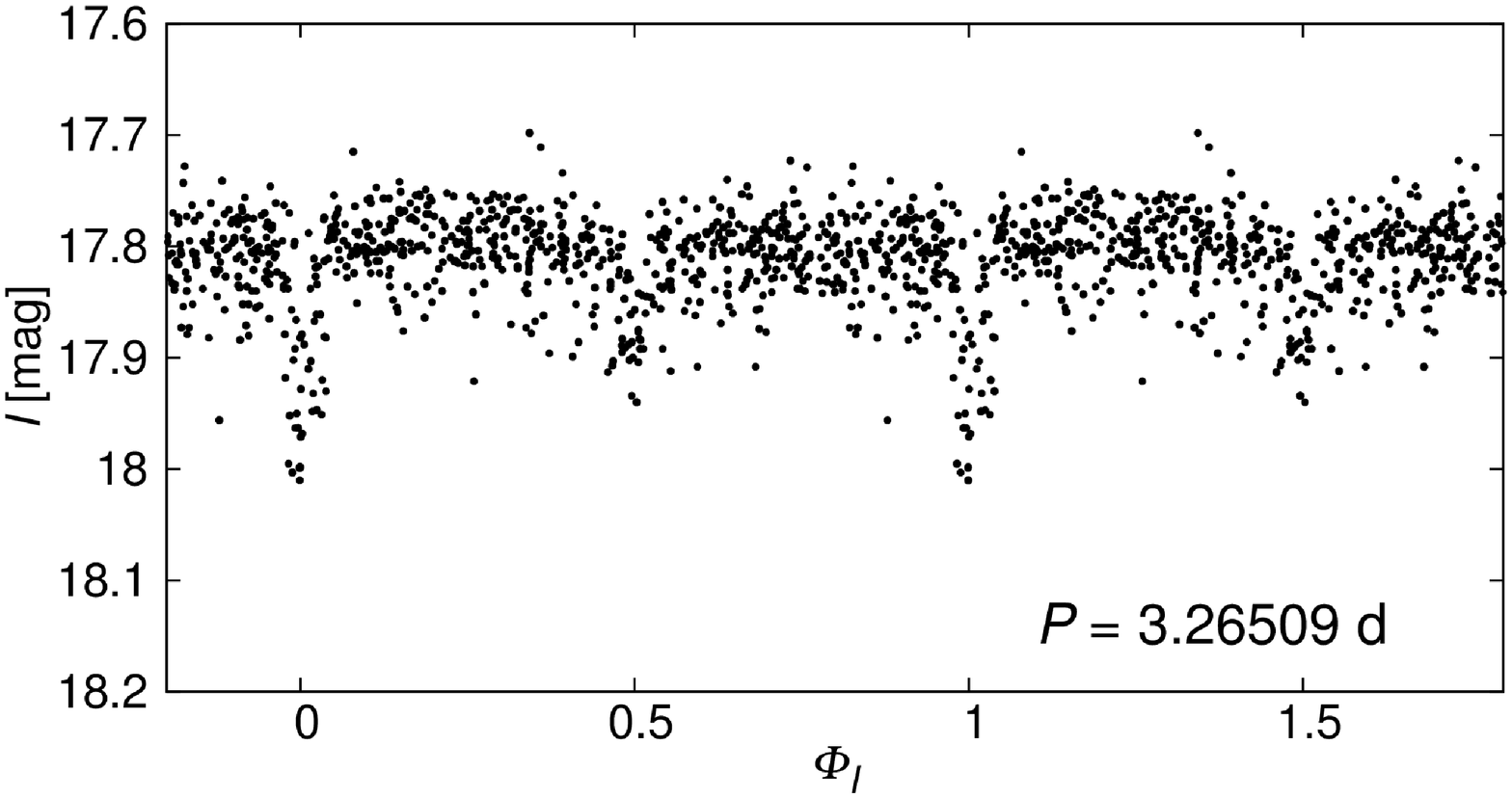} \\
\FigCap{Candidates for double binary systems: OGLE-SMC-ECL-2208 ({\it upper
panels}), OGLE-SMC-ECL-2715 ({\it middle panels}), and OGLE-SMC-ECL-4418
({\it lower panels}). Left and right panels show two separated components.}
\centerline{\includegraphics[angle=270,width=10cm]{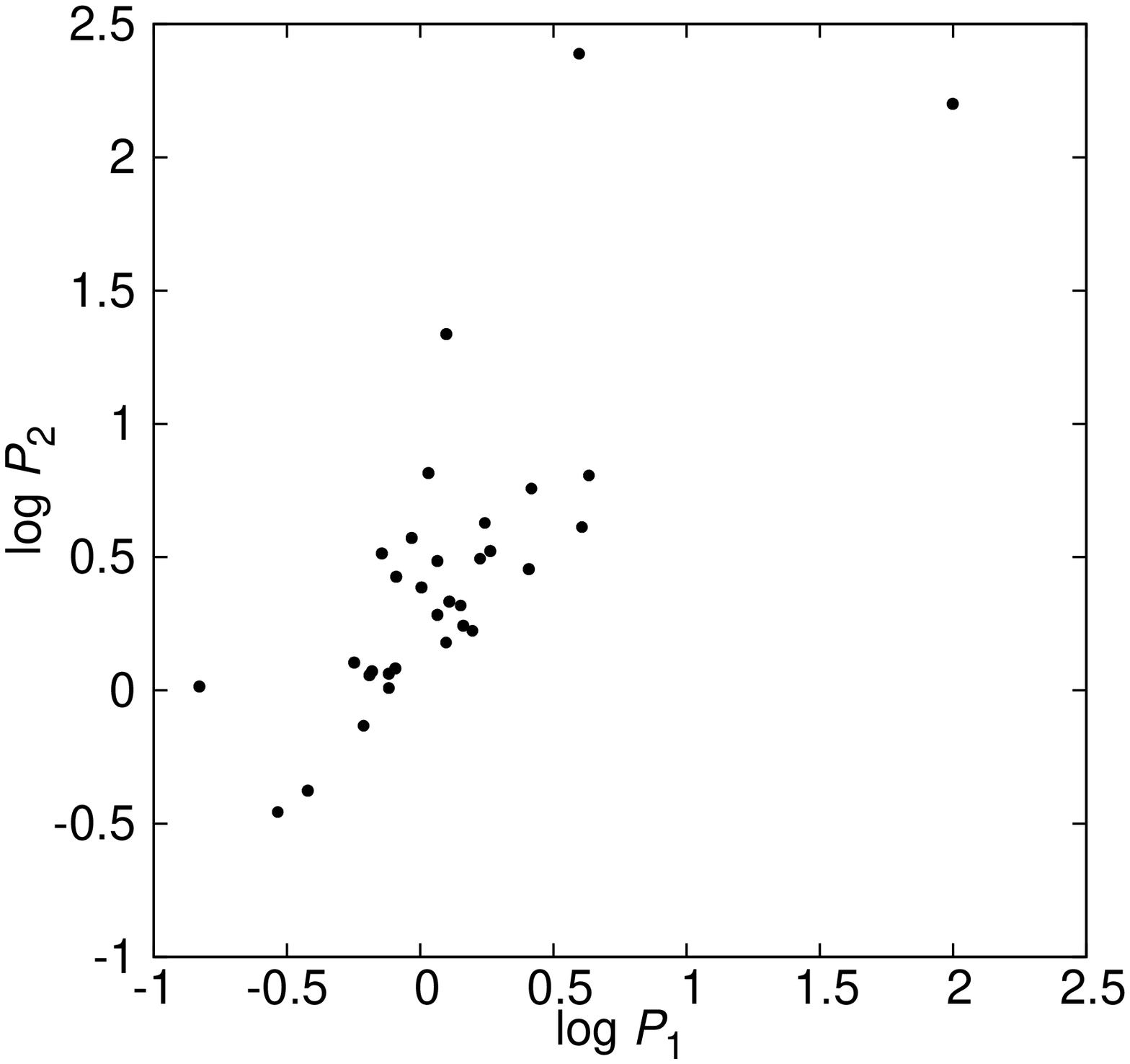}}
\FigCap{Relation between two periods $P_1$ and $P_2$ for all known
candidates for doubly eclipsing quadruple systems found in this paper,
and all previously known ones.}
\end{figure}
A relation between the periods $P_1$ and $P_2$ for all doubly eclipsing
quadruple candidates from the SMC, together with eleven candidates reported
in Pietrukowicz \etal (2013) in the OGLE-III Galactic disk data, and six
previously known Galactic field stars (Batten and Hardie 1965, Harmanec \etal
2007, Lee \etal 2008, Cagas and Pejcha 2012, Lehmann \etal 2012, Lohr \etal
2013) is shown in Fig.~7. There seems to be a weak correlation between
these two parameters.

\subsection{Systems with Noticeable Apsidal Motion}
Our set of eclipsing binaries also contains 32 objects which appear to have
eccentric orbits with a noticeable apsidal motion. In these objects each of
the eclipses can be phased with a slightly different period, so they cannot
be phased together. The most likely explanation of this phenomenon is the
presence of a third body in the system. In Fig.~8, we show an example of
such system -- OGLE-SMC-ECL-0888. The relative differences of periods in
these SMC systems range from $1.53\times10^{-5}$ to $2.19\times10^{-4}$.
\begin{figure}[htb]
OGLE-SMC-ECL-0888\\
\includegraphics[angle=270,width=60mm]{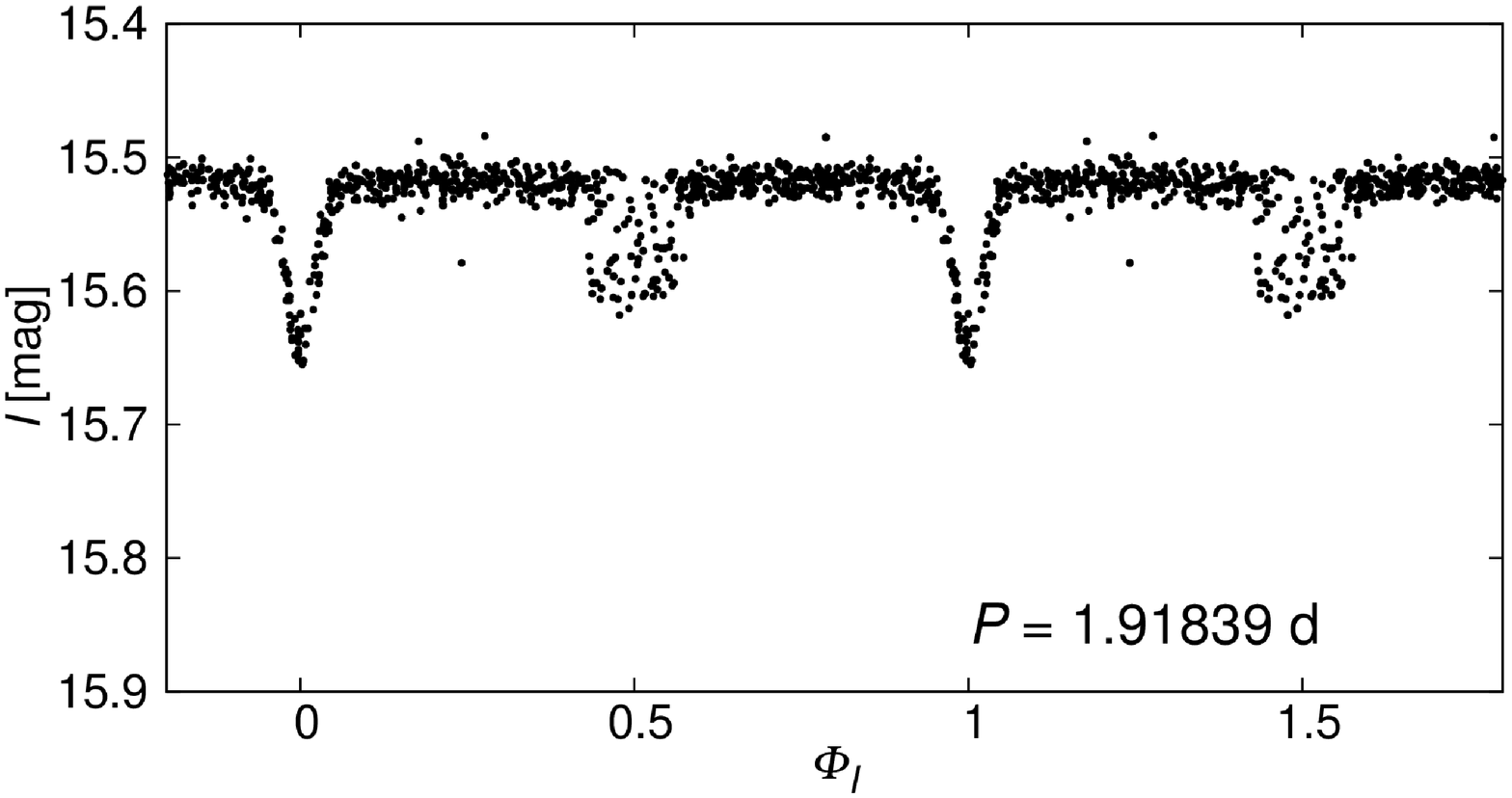} \hfill \includegraphics[angle=270,width=60mm]{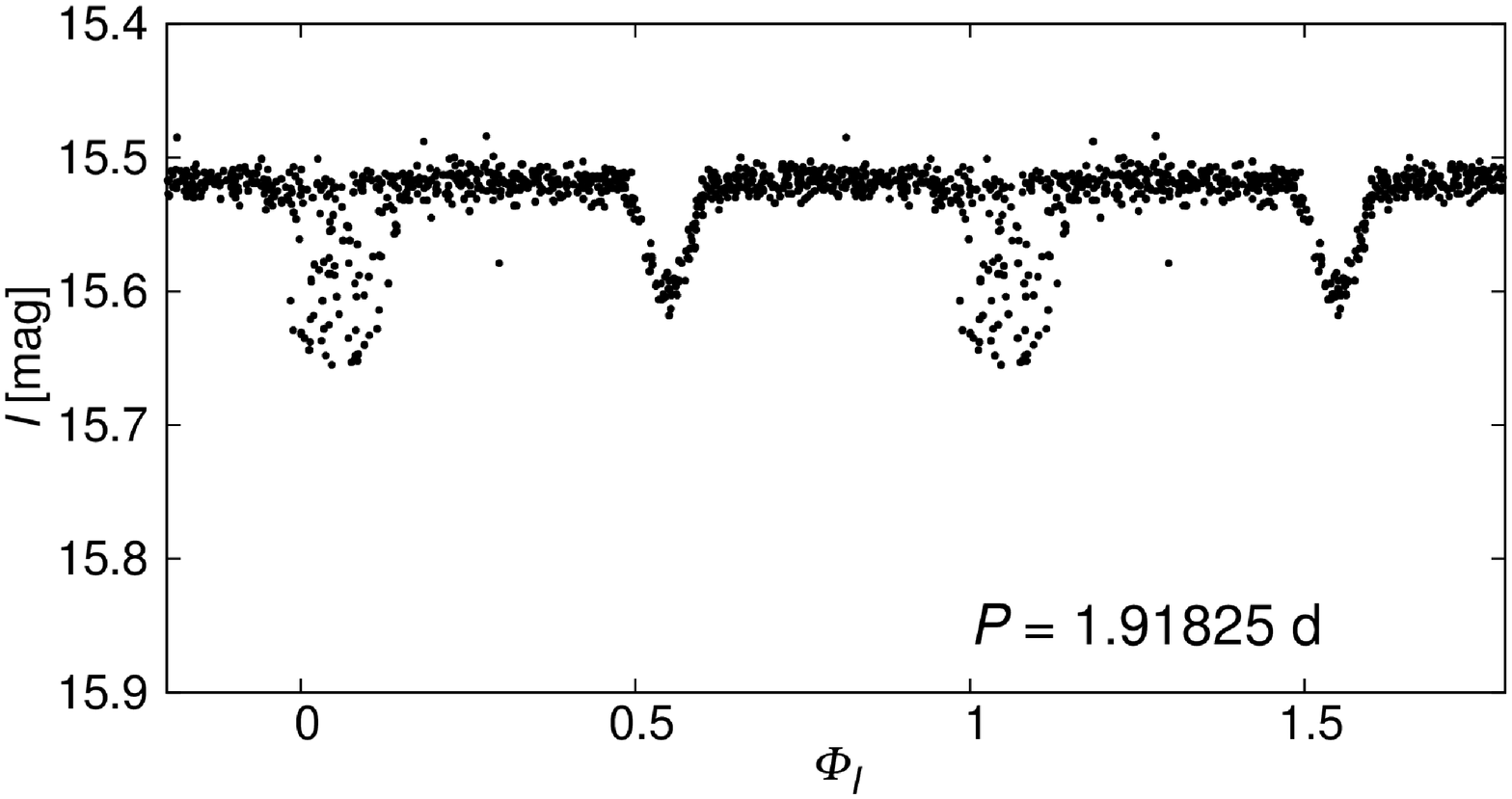} 
\FigCap{Example of a system with eccentric orbit with clear apsidal
motion: OGLE-SMC-ECL-0888. {\it Panels} show the same light curve phased with
two different periods.}
\end{figure}

\subsection{Transient Eclipsing Binary}
\begin{figure}[b]
OGLE-SMC-ECL-5096 \\
\includegraphics[angle=270,width=60mm]{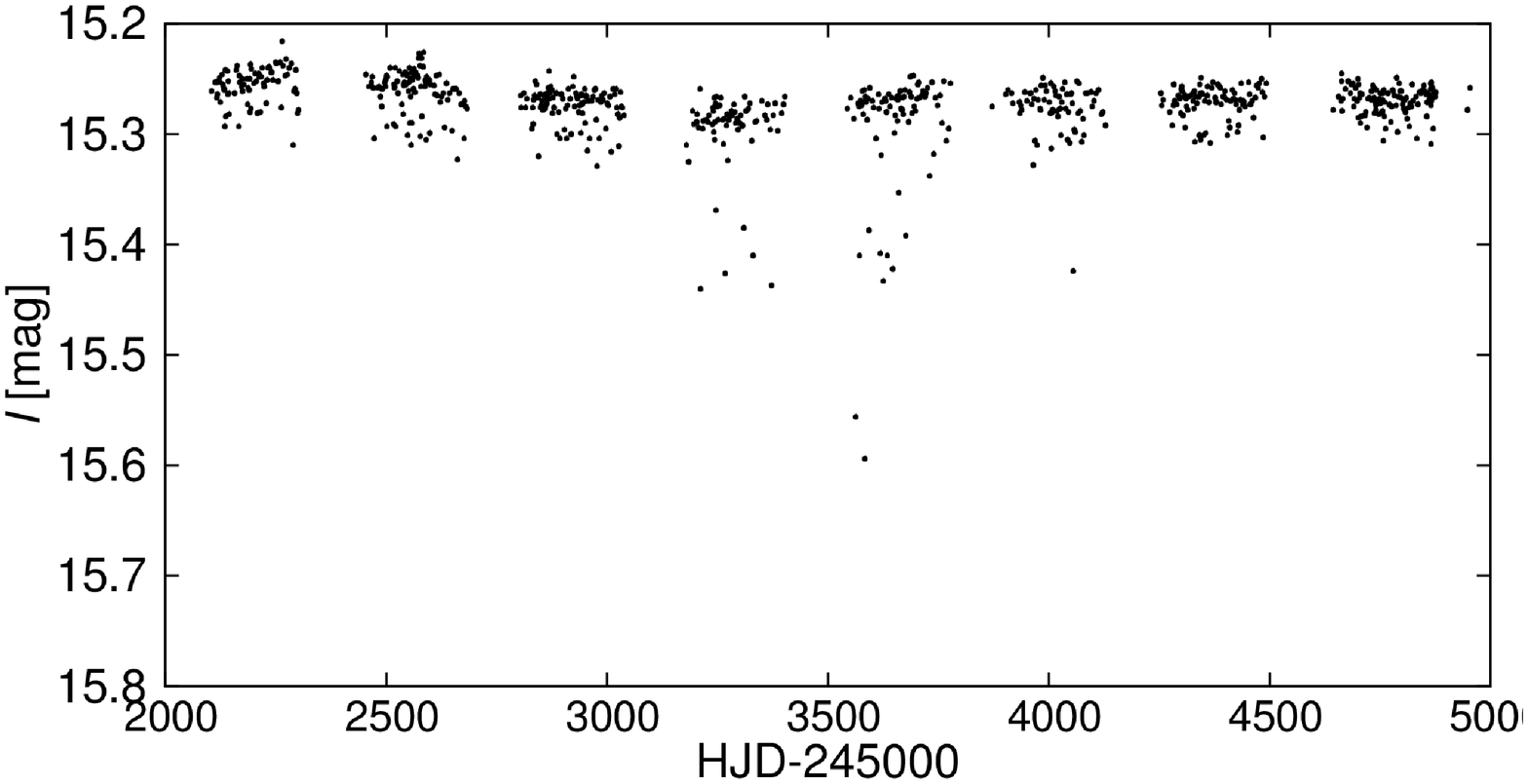} \hfill \includegraphics[angle=270,width=60mm]{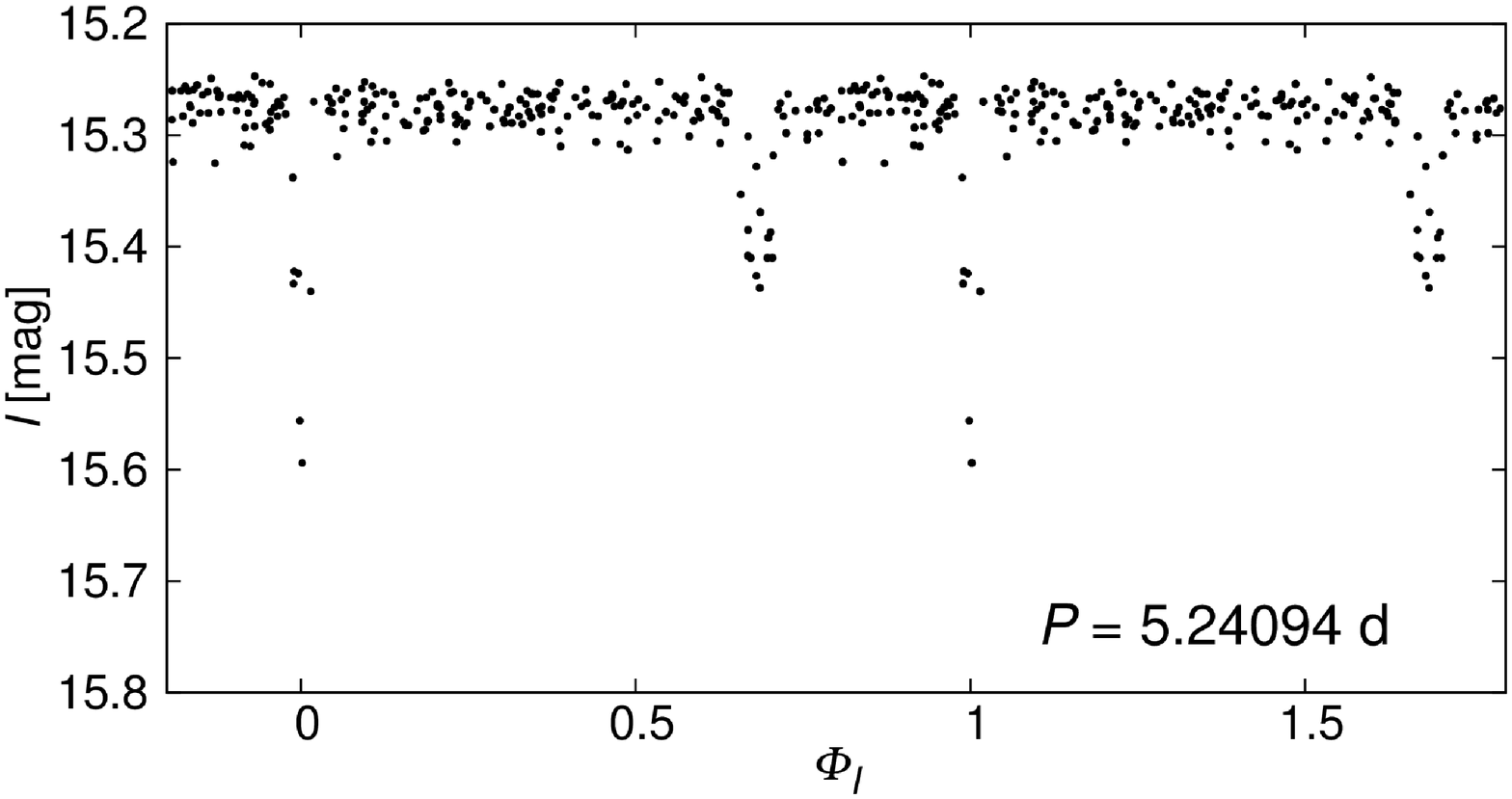}
\FigCap{Transient Eclipsing Binary OGLE-SMC-ECL-5096: unphased light
curve ({\it left panel}) and phased curve for three seasons of observations
(HJD between 2453100 and 2454200 -- {\it right panel}).}
\end{figure}
One of the objects worth noticing is a Transient Eclipsing Binary (TEB)
OGLE-SMC-ECL-5096. Such stars show eclipsing variability only for limited
amount of time. This is due to the precession of the orbital plane or
regression of the nodes (S{\"o}derhjelm 1975). The phased and unphased light
curves of OGLE-SMC-ECL-5096 are presented in Fig.~9. The eclipsing
variability period occurred only during three seasons.

\subsection{RS~CVn Type Stars}
We identified ten eclipsing RS~CVn type stars in our SMC set. These are
chromospherically active close binary stars showing evidences for spots
(Andrews 1998). They can also be sources of X-ray flares (Pandey and Singh
2012) and show signs of accretion (Ró¿yczka \etal 2013). Two of such
systems, presented in Fig.~10, show quite regular sinusoidal variability
apart from the eclipses. The third one shows strong secondary type of
variability, with different period than that of eclipses.
\begin{figure}[h]
OGLE-SMC-ECL-0725 \hglue5cm OGLE-SMC-ECL-6067 \\
\includegraphics[angle=270,width=60mm]{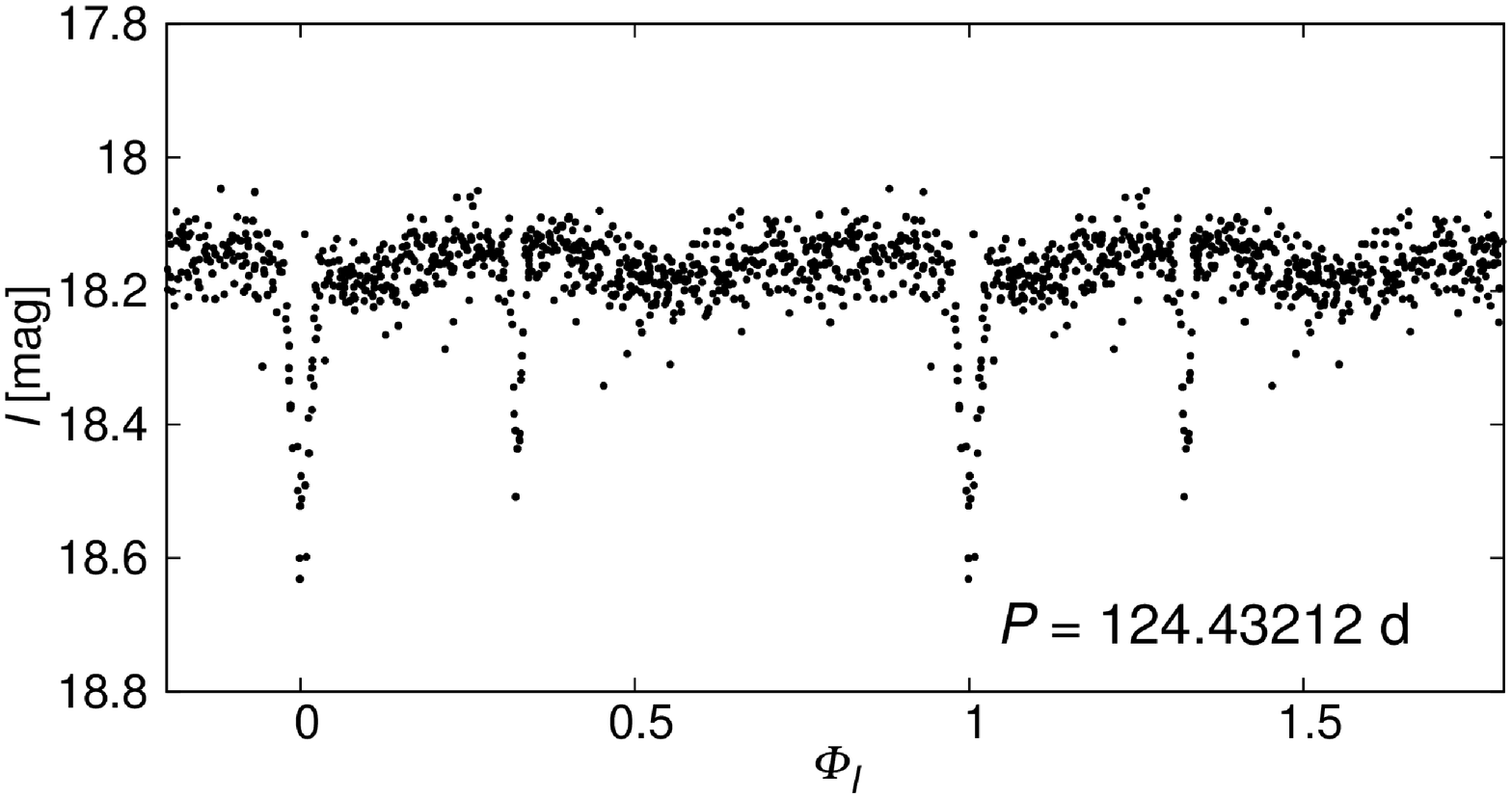} \hfill \includegraphics[angle=270,width=60mm]{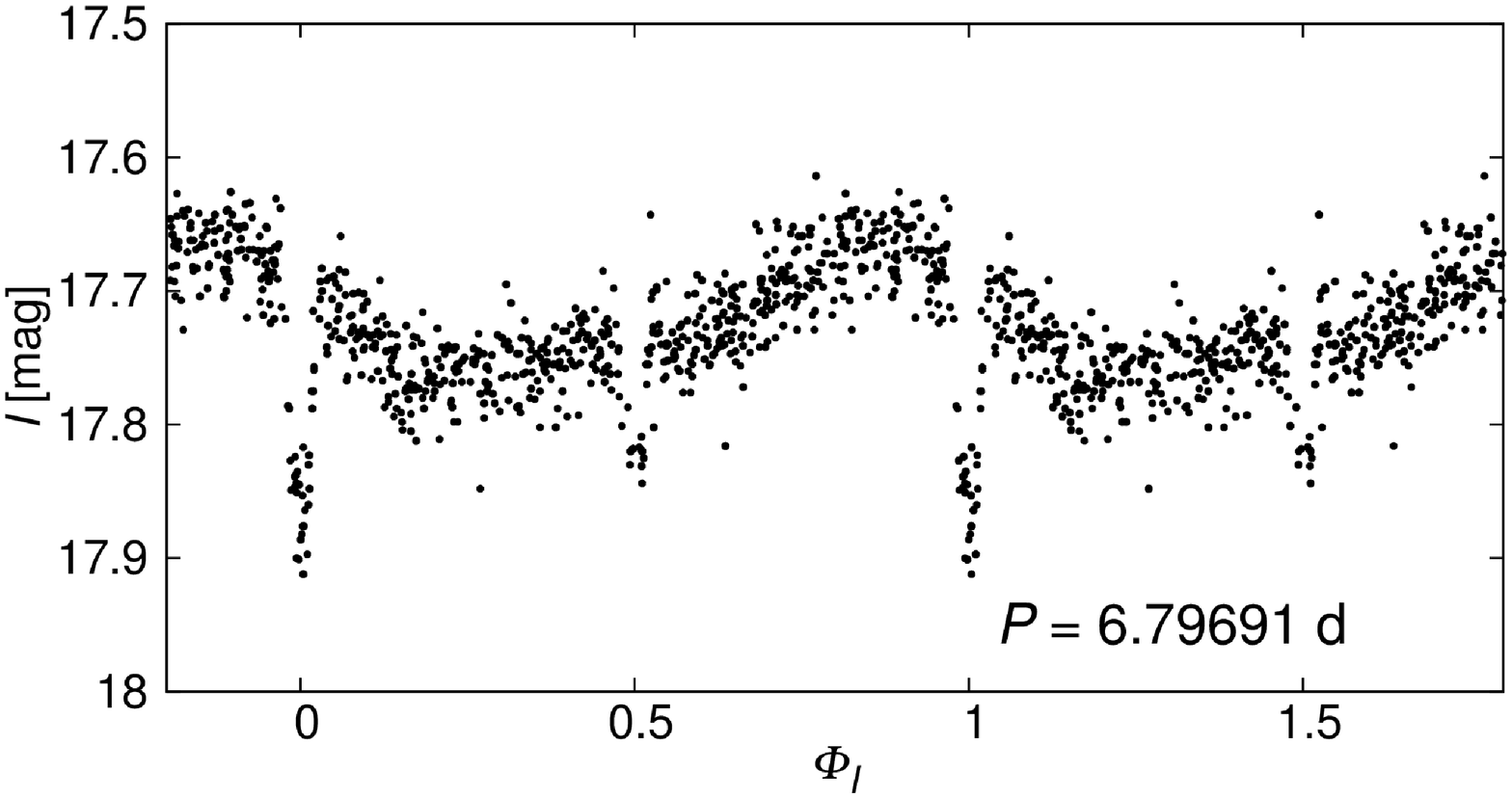} \\ 
OGLE-SMC-ECL-0919 \\
\includegraphics[angle=270,width=60mm]{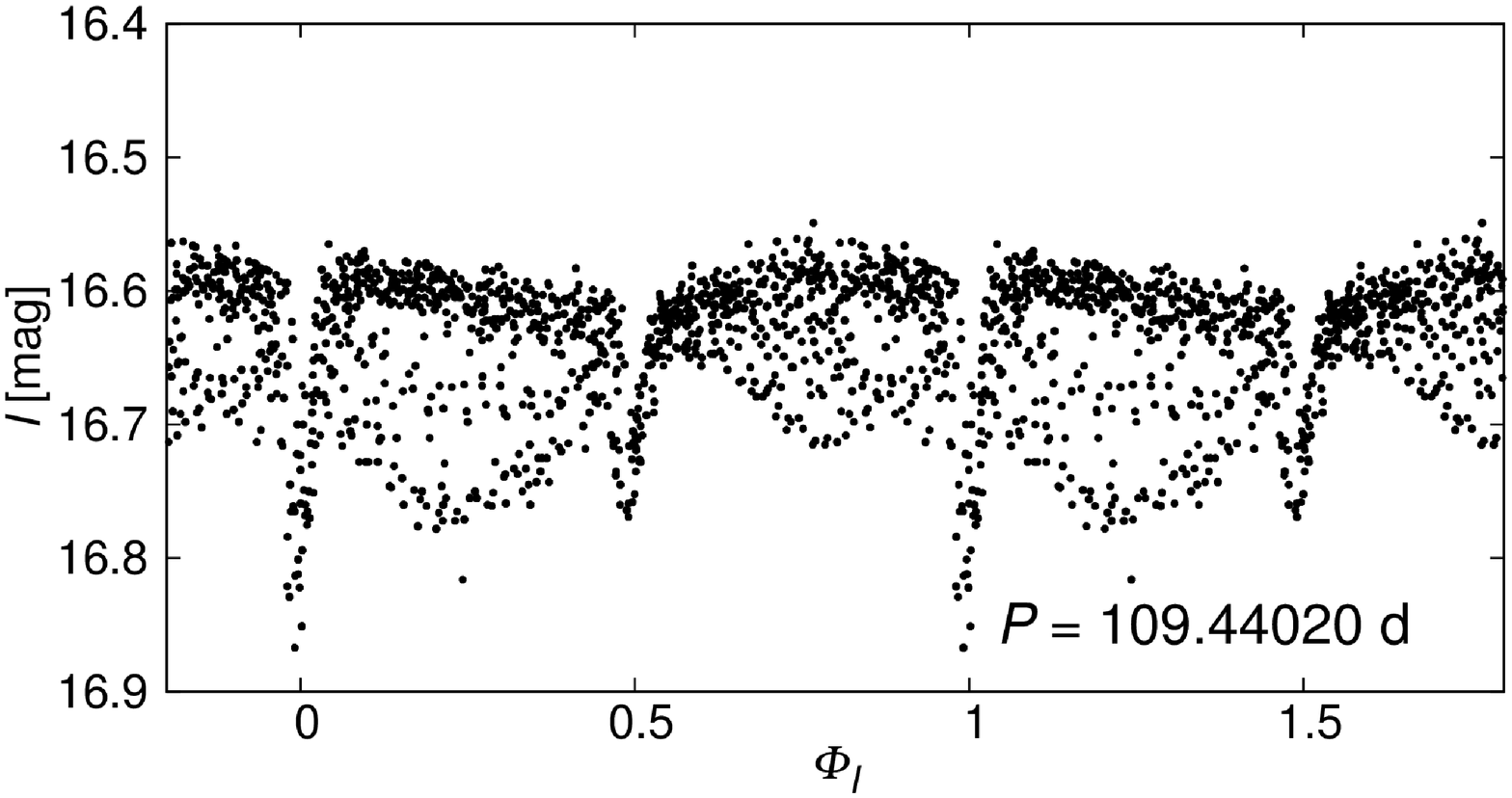}\\ 
\FigCap{RS~CVn type stars: OGLE-SMC-ECL-0725 ({\it upper left panel}),
OGLE-SMC-ECL-6067 ({\it upper right panel}), and OGLE-SMC-ECL-0919 ({\it lower
panel}).}
\end{figure}

\subsection{Double Period Variables}
We found 22 Double Period Variables (DPVs) among our eclipsing
objects. There are also seven stars resembling DPVs but their
classification is uncertain. The characteristic feature of DPVs is the
second type of variability, which has period correlated with the primary
one with the ratio of $P_2/P_1\approx35.2$ (Mennickent \etal 2003). The
second type of variability may be related to relaxation cycles of the
circumprimary disk as suggested by Mennickent and Ko³aczkowski (2010).

For two systems, OGLE-SMC-ECL-1983 and OGLE-SMC-ECL-1830, two maxima in the
long period variability were observed, one significantly higher than the
other, as shown in Fig.~11. Similar light curves were detected by Poleski
\etal (2010) in the OGLE sample of DPVs in the LMC. Such shape of the light
curve can be explained by spots on magnetically active stars (Stêpieñ 1968)
or by spotted young stellar objects (Klagyivik \etal 2013). Thus, these
OGLE DPVs may suggest that the secondary period variability of DPV stars can be
caused by spotted component in the system. It was also suggested by
Mennickent (2012) that DPVs can be progenitors of Be stars.
\begin{figure}[htb]
OGLE-SMC-ECL-1830 \\
\includegraphics[angle=270,width=60mm]{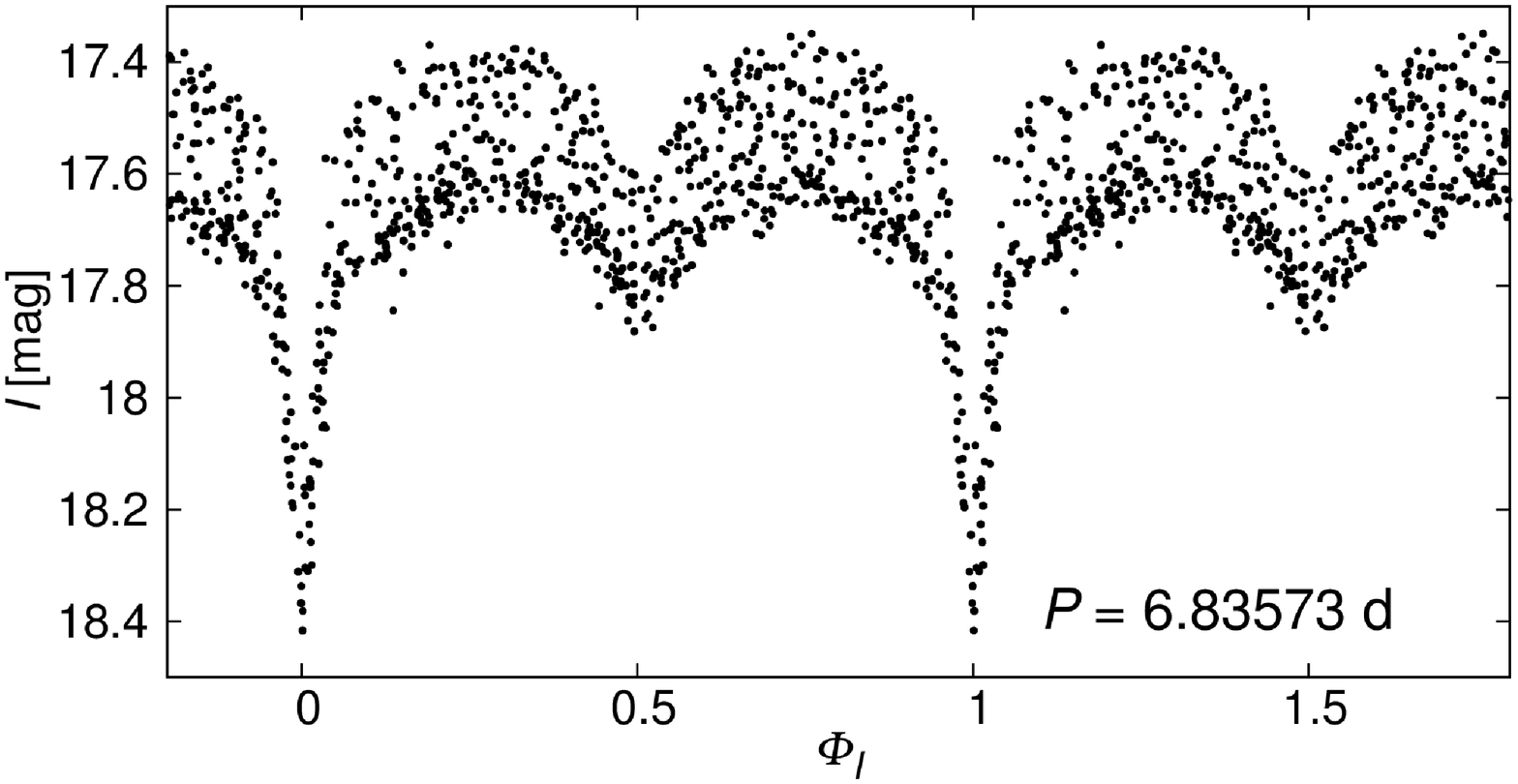} \hfill \includegraphics[angle=270,width=60mm]{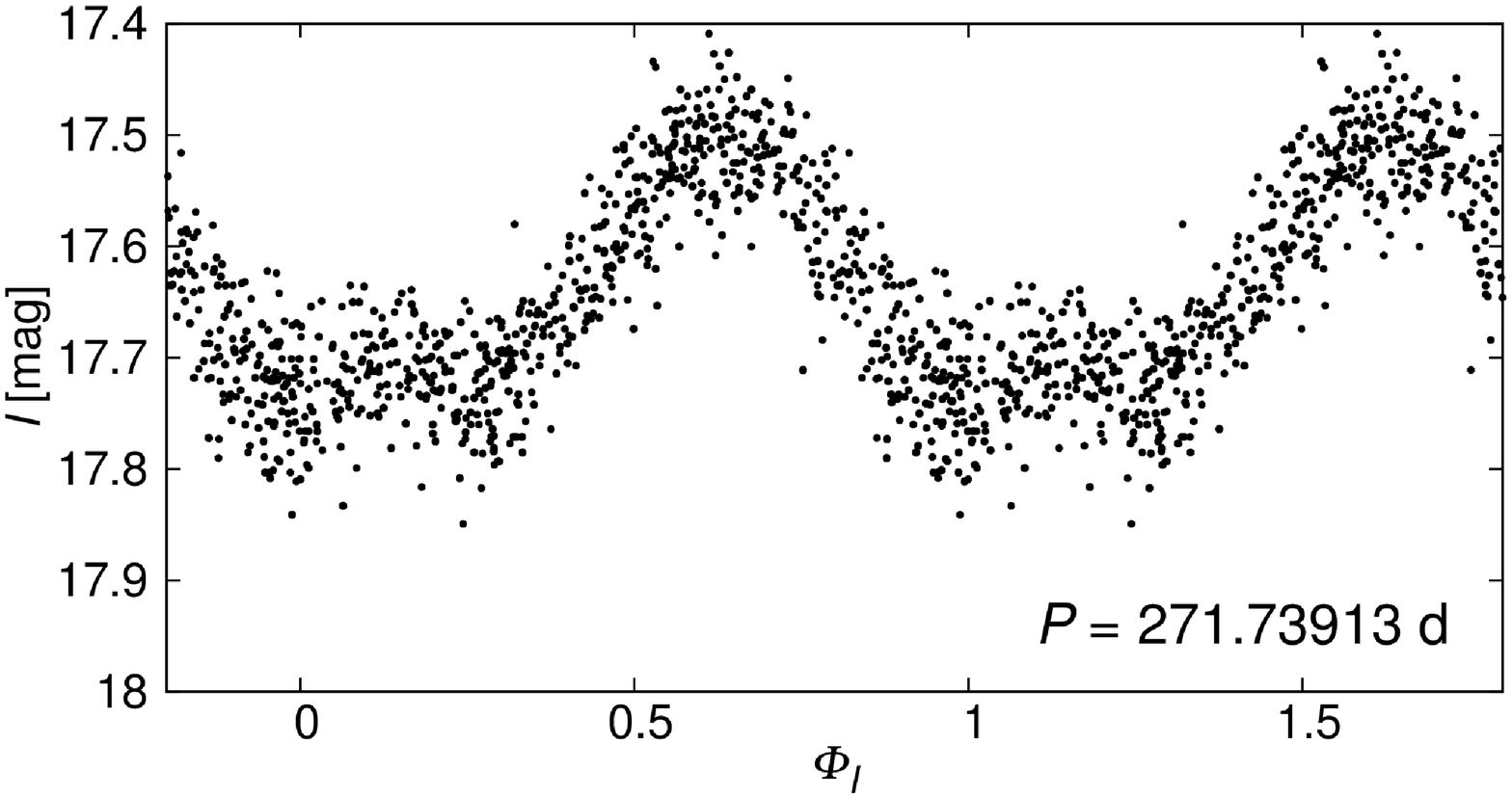} 
\FigCap{DPV type star OGLE-SMC-ECL-1830: the original light curve ({\it left
panel}) and secondary period variability after pre-whitening the curve
with the eclipsing modulation ({\it right panel}).}
\end{figure}

Another interesting example in our sample is the system
OGLE-SMC-ECL-2049. For this star, we can observe that the long period is
changing rapidly during the course of the survey. In Fig.~12, we show the
unphased curve of the second type of variability, as the period change is
too fast to present it phased.
\begin{figure}[htb]
OGLE-SMC-ECL-2049 \\
\includegraphics[angle=270,width=60mm]{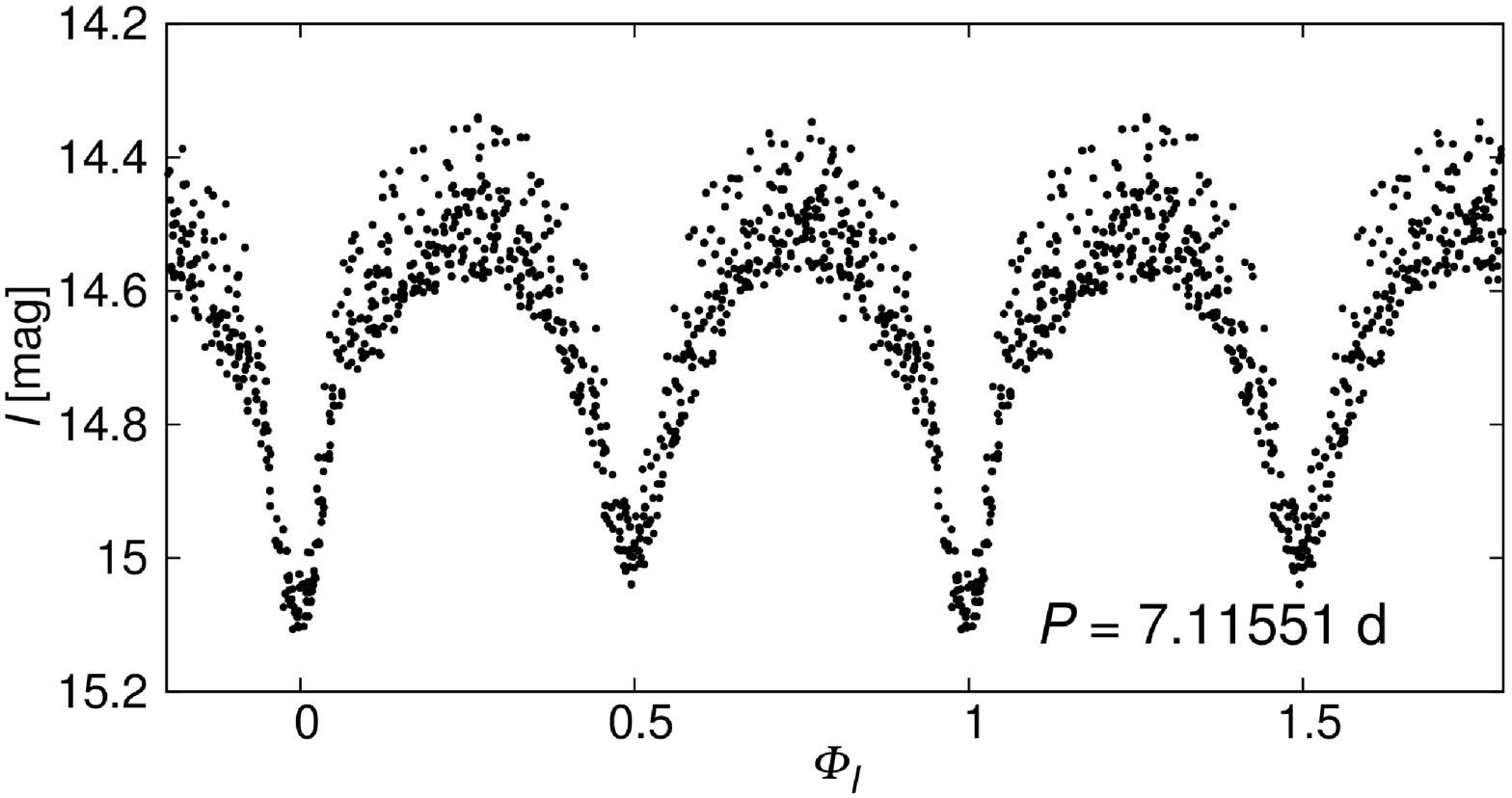} \hfill \includegraphics[angle=270,width=60mm]{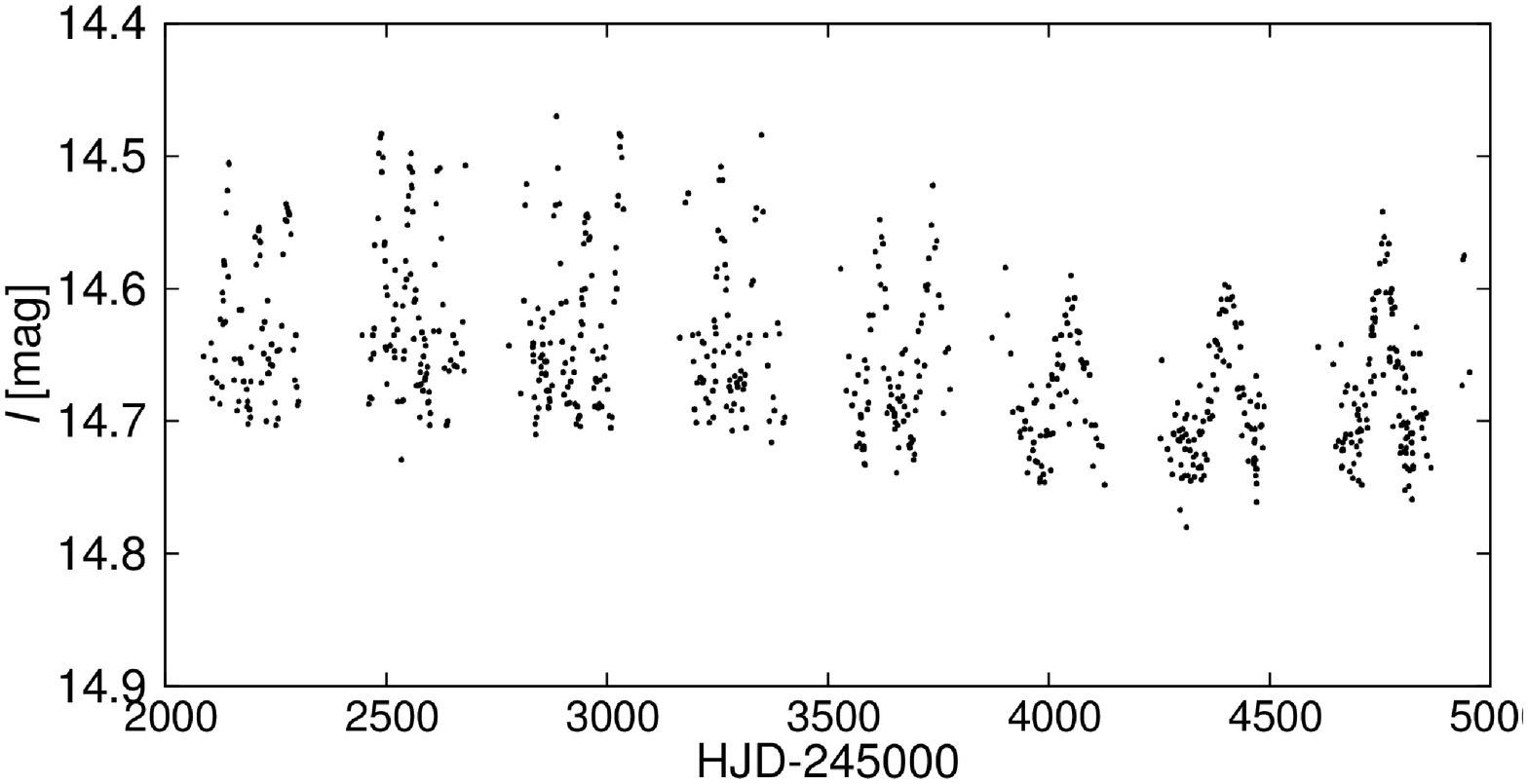} 
\FigCap{DPV type star OGLE-SMC-ECL-2049: the original phased light curve
({\it left panel}) and secondary variability unphased curve after
pre-whitening the eclipsing modulation ({\it right panel}).}
\end{figure}

The list of all found DPVs is presented in Table~2. The system
OGLE-SMC-ECL-2519 has the orbital period $P_1=172.35$~d, which makes it the
longest known DPV. The previous longest one found by Poleski \etal (2010)
has $P_1=109.43$~d. The average value of the periods ratio for the stars in
our sample is $P_2/P_1=31.72$.

\MakeTableee{lrrr}{12.5cm}{List of identified DPVs}
{\hline
\noalign{\vskip3pt}
\multicolumn{1}{c}{ID} & \multicolumn{1}{c}{$P_1$ [d]} & \multicolumn{1}{c}{$P_2$ [d]} & \multicolumn{1}{c}{$P_2 / P_1$} \\
\noalign{\vskip3pt}
\hline
\noalign{\vskip3pt}
OGLE-SMC-ECL-4756 & 0.91773 &  31.16 & 33.9545 \\
OGLE-SMC-ECL-2090 & 3.46263 & 106.13 & 30.6504 \\
OGLE-SMC-ECL-1144 & 4.13872 & 133.73 & 32.3095 \\
OGLE-SMC-ECL-1028 & 4.15963 & 129.08 & 31.0316 \\
OGLE-SMC-ECL-5046 & 4.19228 & 137.92 & 32.8985 \\
OGLE-SMC-ECL-1005 & 4.45509 & 152.37 & 34.2013 \\
OGLE-SMC-ECL-3248 & 4.71304 & 131.45 & 27.8907 \\
OGLE-SMC-ECL-5062 & 4.78243 & 149.01 & 31.1578 \\
OGLE-SMC-ECL-1807 & 4.84151 & 165.95 & 34.2765 \\
OGLE-SMC-ECL-2331 & 4.91588 & 160.00 & 32.5476 \\
OGLE-SMC-ECL-3604 & 5.02385 & 202.01 & 40.2102 \\
OGLE-SMC-ECL-4181 & 5.05442 & 176.14 & 34.8487 \\
OGLE-SMC-ECL-1003 & 5.10733 & 208.14 & 40.7532 \\
OGLE-SMC-ECL-5637 & 5.60217 & 149.01 & 26.5986 \\
OGLE-SMC-ECL-4933 & 6.10996 & 202.01 & 33.0624 \\
OGLE-SMC-ECL-1983 & 6.37713 & 228.73 & 35.8672 \\
OGLE-SMC-ECL-1830 & 6.83573 & 271.82 & 39.7648 \\ 
OGLE-SMC-ECL-2049 & 7.11552 & 174.21 & 24.4831 \\
OGLE-SMC-ECL-1799 & 7.15602 & 223.99 & 31.3009 \\
OGLE-SMC-ECL-1402 & 7.47952 & 286.37 & 38.2872 \\
OGLE-SMC-ECL-2597 & 28.5839~ & 1344.93& 47.0520 \\
OGLE-SMC-ECL-2519 & 172.35~~~~ & $>4000$  & -- \\
\noalign{\vskip3pt}
\hline}
\MakeTable{lccr}{12.5cm}{Candidates for the distance determination -- bright, long-period, detached systems}
{\hline
\noalign{\vskip3pt}
\multicolumn{1}{c}{ID} & \multicolumn{1}{c}{$V$ [mag]} & \multicolumn{1}{c}{$(V-I)$ [mag]} & \multicolumn{1}{c}{$P$ [d]} \\
\noalign{\vskip3pt}
\hline
\noalign{\vskip3pt}
OGLE-SMC-ECL-0019 & 17.921 & 1.014 & 144.01 \\
OGLE-SMC-ECL-0195 & 16.826 & 1.238 & 120.47 \\
OGLE-SMC-ECL-0439 & 18.052 & 0.935 & 279.40 \\
OGLE-SMC-ECL-0470 & 17.168 & 0.562 & 142.07 \\
OGLE-SMC-ECL-0708 & 16.756 & 1.243 & 635.09 \\
OGLE-SMC-ECL-0727 & 18.100 & 1.084 & 316.63 \\
OGLE-SMC-ECL-0970 & 17.937 & 1.109 & 191.58 \\
OGLE-SMC-ECL-1194 & 17.709 & 0.535 & 55.44 \\
OGLE-SMC-ECL-1421 & 17.149 & 0.921 & 102.89 \\
OGLE-SMC-ECL-1492 & 17.699 & 1.074 & 73.76\\
OGLE-SMC-ECL-1567 & 18.849 & 1.191 & 110.87 \\
OGLE-SMC-ECL-1859 & 17.678 & 0.920 & 75.58 \\
OGLE-SMC-ECL-2761 & 17.316 & 0.841 & 150.42 \\
OGLE-SMC-ECL-2841 & 16.931 & 0.764 & 61.46 \\ 
OGLE-SMC-ECL-2876 & 17.484 & 0.887 & 120.93 \\
OGLE-SMC-ECL-3120 & 18.326 & 0.922 & 83.78 \\
OGLE-SMC-ECL-3529 & 17.132 & 0.821 & 234.46 \\
OGLE-SMC-ECL-3678 & 15.531 & 0.718 & 187.97 \\
OGLE-SMC-ECL-4152 & 15.212 & 0.979 & 185.21 \\
OGLE-SMC-ECL-4370 & 18.425 & 0.938 & 63.95 \\ 
OGLE-SMC-ECL-4922 & 16.973 & 1.372 & 1173.19 \\
OGLE-SMC-ECL-5123 & 15.769 & 1.067 & 371.64 \\
OGLE-SMC-ECL-5758 & 18.490 & 1.056 & 100.58 \\
\noalign{\vskip3pt}
\hline }

\subsection{Candidates for the Distance Determination}
Eclipsing binaries can be used as an excellent tool for distance
determination (Pietrzyñski \etal 2013). Eclipsing systems which are
particularly useful for that purpose are bright, red, long-period detached
binaries, with eclipses of comparable depth. Precisely calibrated surface
brightness--color relation for late type stars (currently at 2\% level)
allows basically direct determination of the distance when accurate linear
dimensions of the components are known from eclipsing system solution.

Eclipsing systems from our SMC set which are particularly well suited for
distance determination are listed in Table~3. It is worth noticing that the
distance to the SMC has already been measured by Graczyk \etal (2012, 2013)
based on a few late-type red eclipsing giants discovered by the OGLE
survey.

\Section{Summary and Conclusions}    
In this paper, we report the identification of 6138 eclipsing binary
systems in the Small Magellanic Cloud. This is the largest sample of such
stars in the SMC. We managed to successfully implement the machine learning
technique based on decision tree algorithms for the purpose of the
automatic selection and classification. However, its effectiveness still
needs improvements as it is lower than the statistical approach proposed by
Graczyk and Eyer (2010).

The completeness of our SMC set of eclipsing stars is estimated to be at
the 82\% level, although it depends strongly on the brightness and
amplitude of the stars. The limiting magnitude is $I\approx20.5$~mag, but
the completeness drops fast for objects fainter than 18.5~mag. The systems
were divided into EC and non-EC ones, based on visual analysis of the light
curves. Most of the objects were assigned to the latter group. The EC stars
belong mostly to the main sequence, while the non-EC ones occupy both the
main sequence and the red giant branch.

Fifteen candidates for double binaries, 32 systems with eccentric orbit
having noticeable movement of the line of apsides, a Transient Eclipsing
Binary, ten RS~CVn type stars, and 22 DPV stars were found in our set of
eclipsing binaries in the SMC. In the last group, we identified a system
with a very fast change of the long period, stars with the variability
resembling spotted stars, and a DPV with the longest period known. Finally,
we selected 23 systems, which may be used for the accurate distance
determination to the SMC.

\Acknow{We would like to thank Z. Ko³aczkowski and A. Schwarz\-enberg-Czerny
for providing their period-searching codes. This work has been
supported by the Polish National Science Centre grant No.
DEC-2011/03/B/ST9/02573.

The OGLE project has received funding from the European Research Council
under the European Community's Seventh Framework Programme
(FP7/2007-2013) / ERC grant agreement no. 246678 to AU.}

\end{document}